\documentclass[aps,twocolumn,english,superscriptaddress,citeautoscript,preprintnumbers,amsmath,amssymb,floatfix,footinbib]{revtex4-2}
\usepackage[breaklinks=true,colorlinks,citecolor=blue,linkcolor=blue,urlcolor=blue]{hyperref}
\usepackage{amsmath}
\usepackage{graphicx}
\usepackage{dcolumn}
\usepackage{float}
\usepackage[utf8]{inputenc}
\usepackage[version=4]{mhchem}
\usepackage{color}
\usepackage{soul}
\usepackage[T1]{fontenc}
\usepackage{siunitx}

\begin{document}
	
\title{Quantum oscillations and transport properties of layered single-crystal SrCu$_4$As$_2$}
\author{Sudip Malick}
\email{sudip.malick@pg.edu.pl}
\author{Micha\l{} J. Winiarski}
\email{michal.winiarski@pg.edu.pl}
\affiliation{Faculty of Applied Physics and Mathematics, Gdansk University of Technology, Narutowicza 11/12, 80-233 Gdańsk, Poland}
\affiliation{Advanced Materials Center, Gdansk University of Technology, Narutowicza 11/12, 80-233 Gdańsk, Poland}
\author{Joanna B\l{}awat}
\affiliation {National High Magnetic Field Laboratory, Los Alamos National Laboratory, Los Alamos, NM 87545, USA}
\author{Hanna \ifmmode \acute{S}\else \'{S}\fi{}wi\k{a}tek}
\affiliation{Faculty of Applied Physics and Mathematics, Gdansk University of Technology, Narutowicza 11/12, 80-233 Gdańsk, Poland}
\affiliation{Advanced Materials Center, Gdansk University of Technology, Narutowicza 11/12, 80-233 Gdańsk, Poland}
\author{John Singleton}
\affiliation {National High Magnetic Field Laboratory, Los Alamos National Laboratory, Los Alamos, NM 87545, USA}
\author{Tomasz Klimczuk}
\email{tomasz.klimczuk@pg.edu.pl}
\affiliation{Faculty of Applied Physics and Mathematics, Gdansk University of Technology, Narutowicza 11/12, 80-233 Gdańsk, Poland}
\affiliation{Advanced Materials Center, Gdansk University of Technology, Narutowicza 11/12, 80-233 Gdańsk, Poland}

\begin{abstract}

We report a systematic investigation of the physical properties and Fermi-surface topology of layered single-crystal \ce{SrCu4As2} using electrical transport, magnetotransport, and quantum-oscillation experiments plus band-structure calculations. The temperature-dependent electrical resistivity reveals a hysteretic phase transition at $T_P$ = 59 K, most likely associated with a structural change. Hall resistivity data suggest a marked change in the average hole density resulting from the latter phase transition near $T_P$. A large, linear and nonsaturating magnetoresistance is observed at low temperatures in \ce{SrCu4As2}, likely attributable to the multipocket Fermi surface. Quantum-oscillation data measured in magnetic fields of up to 60 T show several oscillation frequencies exhibiting low effective masses, indicating the presence of Dirac-like band dispersion in \ce{SrCu4As2}, as suggested by the band structure calculations. 

\end{abstract}
	
\maketitle

\section{INTRODUCTION}	

\begin{figure}[h]
	\includegraphics[width=8.5cm, keepaspectratio]{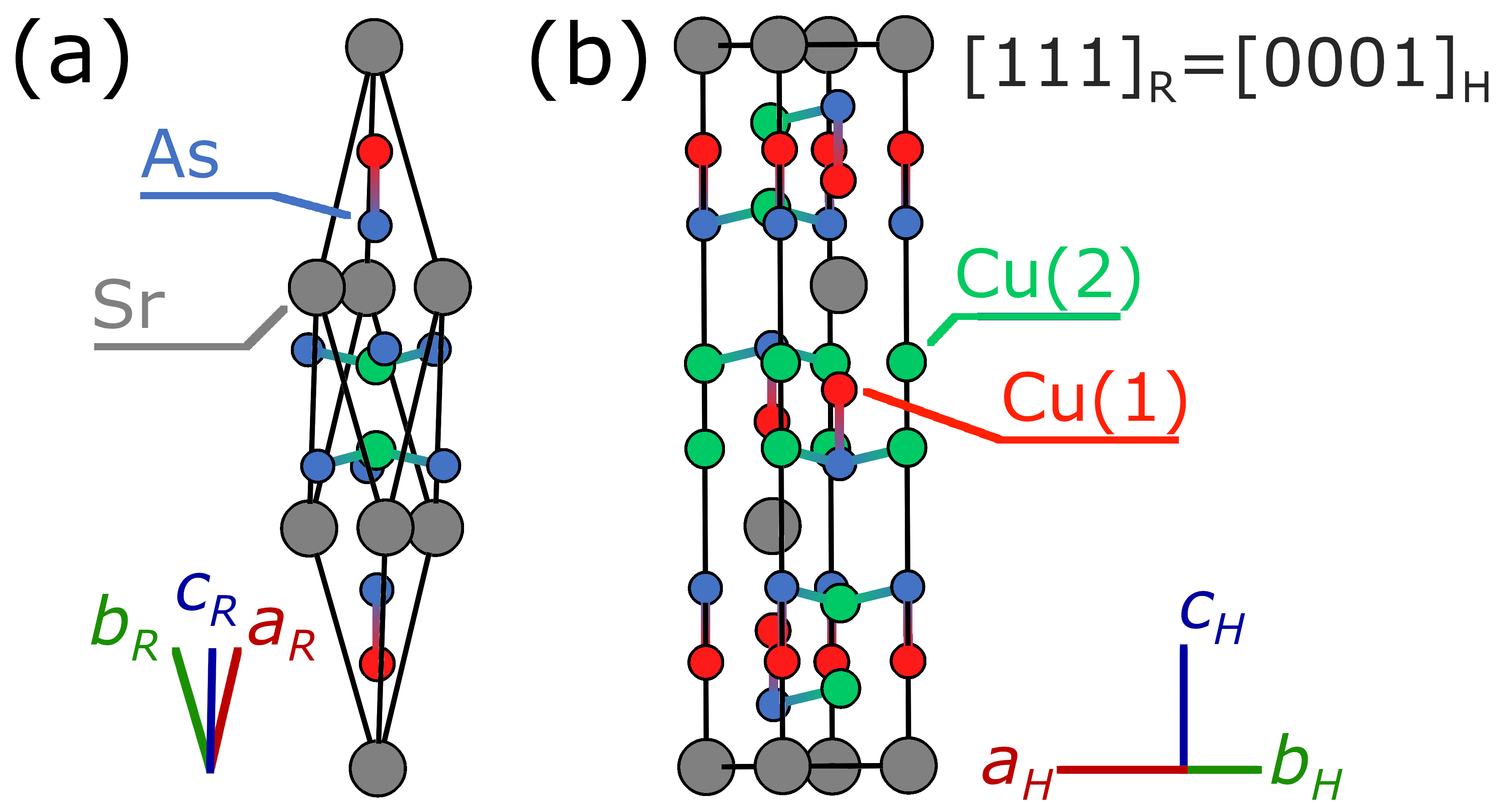}
	\caption{\label{Unitcell} The unit cell of \ce{SrCu4As2} shown using rhombohedral (a) and hexagonal axes (b). There are two symmetry-inequivalent Cu positions in the structure: Cu(1) and Cu(2). Note the [0001] direction in the hexagonal cell is equivalent to the [111] direction in the rhombohedral one. }
\end{figure}

Ternary pnictides with layered crystal structures have drawn significant research interest due to their distinctive physical properties resulting from the interplay of topology, magnetism, superconductivity, charge-density waves (CDWs), and structural disorder \cite{LaAuSb2_2020,LaAuSb2_2023,LaAgSb2_2016,LaAgSb2, PrAgBi2,BaAg2As2, BaFe2As2, BaFe2As2_2011, EuAg4Sb2, CaCu4As2}. \ce{CaCu4As2}-type layered pnictide is one of the relevant systems where large linear magnetoresistance, quantum oscillations, structural phase transition, and unusual magnetism are observed due to their complex Fermi surface topology. Very recently, a spin-moiré superlattice has been realized in layered pnictide \ce{EuAg4Sb2}, establishing the layered "142" system as a potential platform for emerging spin-driven quantum-Hall states \cite{EuAg4Sb2_sciadv}. Our recent report \cite{EuAg4Sb2} indicates complex magnetism and unusual magnetoresistance (MR) in \ce{EuAg4Sb2}. The latter compound features two successive antiferromagnetic phase transitions, multiple metamagnetic transitions, and large nonsaturating MR with quantum oscillations. The presence of several oscillation frequencies exhibiting low effective masses in this compound indicates a complex Fermi surface \cite{EuAg4Sb2}. Interestingly, a structural distortion at 120 K and an incommensurate antiferromagnetic phase transition below 9 K are observed in the arsenide \ce{EuAg4As2} \cite{EuAg4As2_2020,EuAg4As2_2021}.  Similar structural distortion and ferromagnetic phase transition are seen at 75 K and 35 K, respectively, in  isostructural \ce{EuCu4As2} \cite{EuCu4As2}. Moreover, both compounds show large magnetoresistance.  These behaviors suggest that the interplay between magnetism and structural changes plays a significant role in the electronic properties of these materials. While the nonmagnetic arsenides of these series show similar high-temperature structural distortion and large magnetoresistance, they show some additional features. For instance, quantum oscillations linked with small Fermi pockets and a significant change in the Fermi surface topology, in contrast to the band structure calculation, are observed due to structural distortion in \ce{SrAg4As2} \cite{SrAg4As2}. On the other hand, \ce{CaCu4As2} exhibits Shubnikov-de Haas (SdH) oscillations and coexistence of the multilayer quantum-Hall effect and a CDW state  \cite{CaCu4As2}. These findings motivate us to investigate the nonmagnetic analog of \ce{EuCu4As2}, \ce{SrCu4As2}, which crystallizes in a rhombohedral structure (Fig. \ref{Unitcell}) with space group $R\Bar{3} m$ \cite{SrCu4As2_1999}. A recent report on thermal expansion and electrical resistivity data reveals a structural distortion in \ce{SrCu4As2} at 145 K \cite{ACu4As2_2023}. Density-functional-theory calculations indicate that \ce{SrCu4As2} is a semimetal, with highly dispersive bands near the Fermi energy and holelike Fermi sheets along the $\Gamma-Z$ direction \cite{SrCu4AS2_DFT}. However, few thorough investigation of the physical properties of this compound have been carried out. Here, we present a comprehensive study of the physical properties and Fermi-surface topology of high-quality single-crystal \ce{SrCu4As2} using electrical resistivity, magnetotransport, and quantum-oscillation measurements at high magnetic fields along with band-structure calculations.  Our findings indicate that \ce{SrCu4As2} undergoes a possible structural-distortion-type phase transition. Magnetotransport measurements show a linear MR and Hall resistivity. The quantum oscillation reveals a complex Fermi surface with some quasi-two-dimensional sections, in qualitative agreement with our theoretical calculations. The Supplemental Material (SM) contains information on single-crystal growth, structural characterization, measurement techniques, band-structure calculations, and fast Fourier transform (FFT) spectra \cite{SM} (see also Refs. \cite{PDO_Exp, CeOs4Sb12, elk, Model_77, KAWAMURA2019197, ROURKE2012324, ACu4As2_2023} therein).

\section{Results and discussion}
\subsection{Transport properties}

\begin{figure*}
	\includegraphics[width=16.5cm, keepaspectratio]{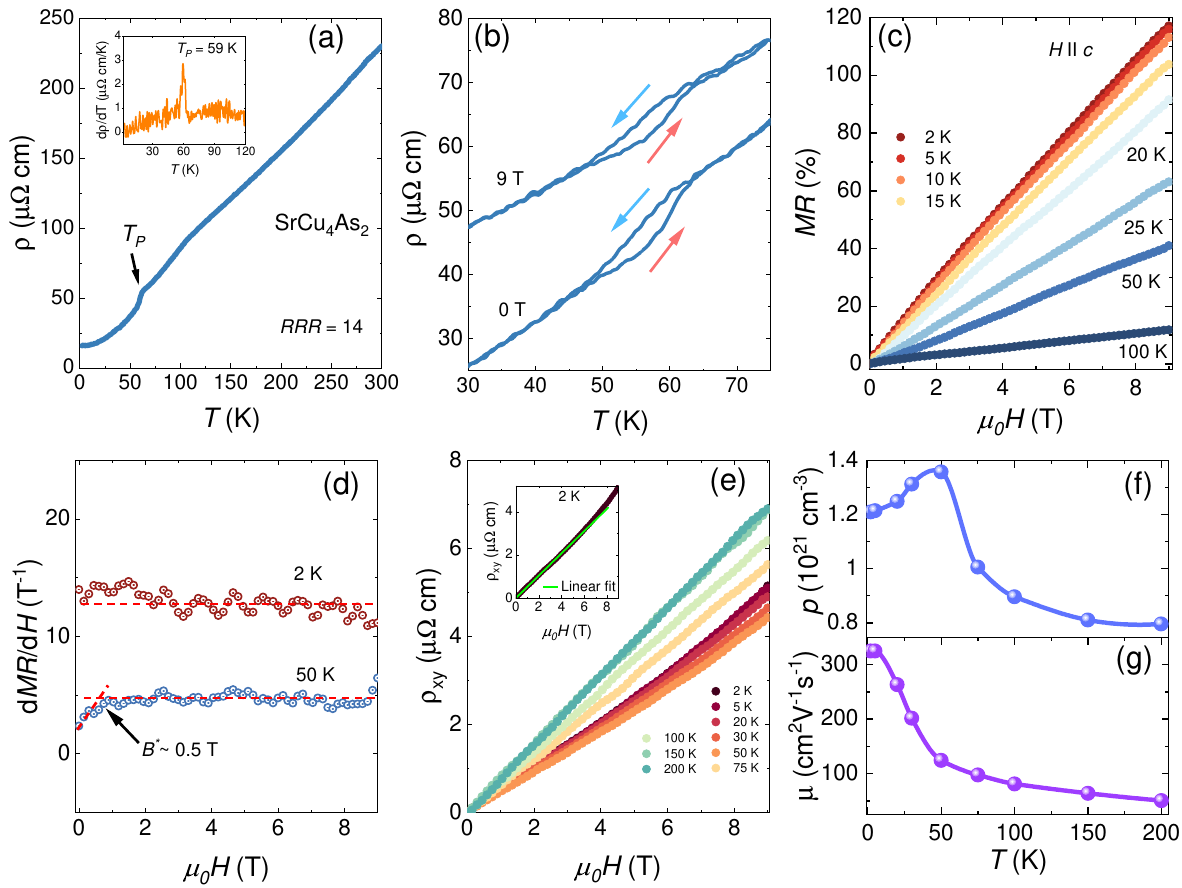}
	\caption{\label{RT} Electrical resistivity and magnetotransport properties of \ce{SrCu4As2}: (a) The temperature dependence of the zero-field electrical resistivity from 2 to 300 K with the current applied in the ab plane of the crystal. The inset shows the first-order derivative of the resistivity data, the peak occurring at the phase transition. (b) Cooling and heating curves for the electrical resistivity near the phase transition temperature at 0 and 9 T. (c) Field dependence of the magnetoresistance measured up to 9 T for various temperatures ($H \parallel c$). (d) The first-order derivative of the MR as a function of magnetic field for temperatures of 2 K and 50 K. (e) The field-dependent Hall resistivity measured between 2 and 200 K. The inset shows the linear fit up to 6 T at 2 K. (f) and (g) display the temperature variation of the hole concentration and the average mobility estimated from the Hall resistivity data. }
\end{figure*}

Figure \ref{RT}(a) shows the zero-field electrical resistivity $\rho$(T) of a \ce{SrCu4As2} single crystal measured over the temperature ($T$) range 2-300 K. The resistivity gradually decreases with falling temperature, before showing a sudden drop at around 60 K. The overall electrical resistivity implies metallic or semimetallic behavior and a residual resistivity ratio (RRR) of 14, which is comparable to isostructural compounds, suggesting that the as-grown crystals are of high quality \cite{CaCu4As2, SrAg4As2}. Note that temperature-dependent heat-capacity data also exhibit a feature at around 60 K  [see the SM \cite{SM}]. The first-order derivative of $\rho$(T) (Fig. \ref{RT}(a) inset) indicates that the phase transition occurs at $T_P$ = 59 K, much lower than the 145 K reported previously \cite{ACu4As2_2023}. Hysteresis is observed in the resistivity data for heating and cooling [Fig. \ref{RT}(b)]; although a significant enhancement of the electrical resistivity is observed due to applied magnetic field, a field of 9 T has no apparent effect on the transition temperature or the hysteresis loop. Similar nonmagnetic phase transitions are also seen in other "142"-type layered compounds \cite{CaCu4As2, SrAg4As2,SrCu4xP2, EuAg4As2_2020,EuCu4As2}. In \ce{SrCu4As2}, the observed phase transition is unlikely to be due to a CDW, as these typically result in an upward jump in electrical resistivity due to the opening up of a partial gap in the Fermi energy \cite{CDW_Rev}. Hence, the high-temperature nonmagnetic phase transition in "142" systems is likely to be a structural distortion \cite{CaCu4As2, SrAg4As2,SrCu4xP2, EuAg4As2_2020,EuCu4As2}. To strengthen this assertion for \ce{SrCu4As2}, future temperature-dependent single-crystal x-ray diffraction would be very helpful.

Figure. \ref{RT}(c) shows the magnetoresistance measured at various temperatures; the magnetic field is applied along the $c$-axis and the current is in the $ab$-plane of the \ce{SrCu4As2} single crystal. With increasing magnetic field, the MR increases linearly and reaches a value of 120\% at 9 T and 2 K. Similar MR values were recently reported in the isostructural compounds \ce{SrAg4As2} \cite{SrAg4As2}, \ce{CaCu4As2} \cite{CaCu4As2}, and \ce{SrCu4P2} \cite{SrCu4xP2}. However, the sizes of these MRs are less than those of typical topological materials \cite{Rev_TM, WTe2}. A substantial decrease in MR is observed as the temperature increases; however, there seems to be no approach to saturation, even at 100 K and 9 T. At low temperatures, \ce{SrCu4As2} exhibits almost linear MR at low temperatures, as shown by the almost field-independent first-order derivative of the MR data at 2 K [Fig. \ref{RT}(d)]. Linear MR is often observed in topological semimetals and insulators, attributed to their linear band dispersions and topologically protected surface states \cite{Rev_TM,TlBiSSe,FeP, YPdBi, Ru2Sn3}. Moreover, some Dirac semimetals exhibit a MR crossover from a weak-field quadratic field dependence to high-field behavior. This has been interpreted using Abrikosov’s quantum-limit theorem, which invokes Dirac carriers being restricted to the zeroth Landau level \cite{BaFe2As2_2011,PrAgBi2,Abrikosov,Tranport_topo}. Whilst there appears to be a small region of quadratic MR in \ce{SrCu4As2}  below $B^* \approx 0.5$ T at $T$ = 50 K, there is no evidence for a similar dependence at 2 K [Figs. \ref{RT}(c) and \ref{RT}(d)]. As we will see below, the quantum oscillations have frequencies in the approximate range 200 to 2000 T; hence, the crossover field observed at 50 K is ~two orders of magnitude below the quantum limit. This indicates that Abrikosov’s theory cannot explain the MR of \ce{SrCu4As2}. The carrier compensation and mobility fluctuations could be the other possible reasons for displaying large MR. The compounds having an equal density of electrons and holes, such as LaSb \cite{LaSb} and \ce{TaAs2} \cite{TaAs2}, exhibit high unsaturated MR due to carrier compensation. Our Hall resistivity data (see below) suggest a dominating hole-type carrier, which eliminates the possibility of carrier compensation in \ce{SrCu4As2}. On the other hand, the observed hole mobility is low compared to compounds like disordered Ag$_{2+\delta}$Se and $n$-doped \ce{Cd3As2}, where mobility fluctuation induces large and linear MR \cite{Ag2Se, Cd3As2, n-Cd3As2}. The linear and unsaturated MR in \ce{SrCu4As2} is likely associated with its multiband Fermi surface with several tiny Fermi pockets, characterized in band-structure calculations by Dirac-like band dispersion near the Fermi energy and small effective masses obtained from the quantum oscillation data (see below).

Figure. \ref{RT}(e) displays the Hall resistivity $\rho_{xy}$ as a function of applied magnetic field measured at various temperatures. The Hall resistivity exhibits a linear field dependence with positive slope for all measured temperatures between 2 and 200 K, suggesting that transport in \ce{SrCu4As2} is dominated by holes. The average carrier concentration ($p$) and Hall mobility ($\mu$) have been calculated using the expressions $p = 1/(eR_H)$ and $\mu = R_H/\rho(H = 0)$, where $R_H$ is the slope of the linear fit to the Hall resistivity data presented in the inset of Fig. \ref{RT}(e). The temperature variation of the average hole concentration and mobility are presented in Figs. \ref{RT}(f) and \ref{RT}(g), respectively. These values are comparable to those of isostructural compounds \ce{SrAg4As2} \cite{SrAg4As2} and \ce{CaCu4As2} \cite{CaCu4As2}.  As \ce{SrCu4As2} passes through the phase transition at 59 K, a sudden change in the average hole density concentration is observed [Fig. \ref{RT}(g)]. Moreover, the hole concentration is a few orders of magnitude lower than typical metals but comparable to several topological semimetals \cite{YbCdGe, ZrSiSe, ZrSiS}, which indicates a semimetallic character of \ce{SrCu4As2}.

\subsection{Quantum oscillations}
\begin{figure*}
	\includegraphics[width=17cm, keepaspectratio]{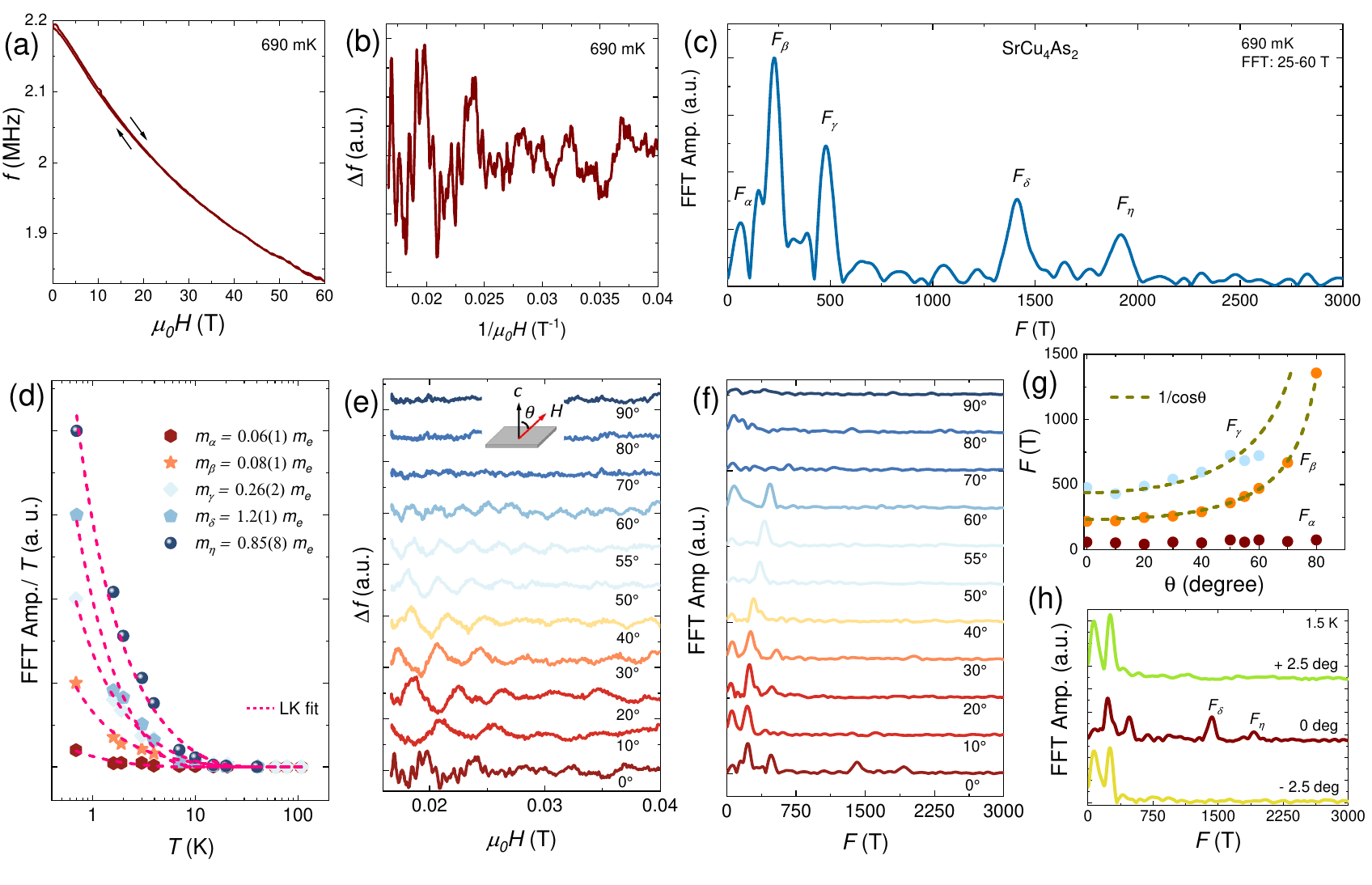}
	\caption{\label{PDO} Quantum oscillations in \ce{SrCu4As2} observed via PDO measurements: (a) The field dependence of PDO frequency as a function for rising and falling fields at 690 mK. (b) PDO data as a function of 1/$\mu_0H$ after polynomial background subtraction in the field range 25-60 T at 690 mK for $H\parallel c$. (c) FFTs of background subtracted PDO data at 690 mK. (d) The temperature variation of the FFT amplitudes fitted using Eq. \ref{LK}. (e) Quantum oscillations at 690 mK for various angles of the magnetic field with respect to the c-axis of the crystal, as shown in the inset diagram. The corresponding FFTs are shown in (f). (g) The angular dependences of the FFT frequencies. The dotted curves represent 1/$\cos{\theta}$ fits. (h) FFTs at 1.5 K for 0$^\circ$ and ±2.5$^\circ$}.
\end{figure*}

Quantum oscillations in \ce{SrCu4As2} were measured using the proximity-detector oscillator technique in a 60 T pulsed magnet and $^3$He cryostat at NHMFL Los Alamos. The PDO circuit’s resonant frequency $f$ is sensitive to the inductance of the coil around the sample, which is determined by the sample’s skin depth. Skin depth is in turn dependent on electrical conductivity; hence Shubnikov-de Haas oscillations are observed as variations in $f$ \cite{PDO_Exp}. Fig. \ref{PDO} (a) shows the PDO signal as a function of magnetic field. A third-order in $H$ polynomial background was subtracted from the PDO signal over the field range 25-60 T; the result is plotted as a function of $1/\mu_0H$ in Fig. \ref{PDO}(b). Fig. \ref{PDO}(c) shows the corresponding fast Fourier transform, revealing five fundamental frequencies: $F_\alpha$ = 55 T, $F_\beta$ = 225 T, $F_\gamma$ = 478 T, $F_\delta$ = 1410 T, and $F_\eta$ = 1914 T. Subsequently, the effective mass associated with each frequency is estimated by fitting the FFT amplitudes to the temperature-damping component of the Lifshits-Kosevich (LK) equation, as given below \cite{Shoenberg}:

\begin{equation}
    \frac{A_{FFT}}{T} \propto \dfrac{14.69m^*/B_m}{\text{sinh}(14.69m^*T/B_m)}~~.
    \label{LK}
\end{equation}

\noindent Here $m^*$ is the effective mass and $B_m$ is calculated using the expression $1/B_m = 2(1/B_l+1/B_u)$, where $B_l$ and $B_u$ are, respectively, the lower and upper field limits of the Fourier window. Since the frequencies $F_\delta$ and $F_\eta$ are more prominent at high fields, a 35-59 T Fourier window was used to derive their effective masses, whereas a 20-60 T Fourier window was used for the lower-frequency mass fits (Quantum oscillations and corresponding FFTs for these two windows can be found in the SM \cite{SM}.). The variation of the FFT amplitudes with temperatures, along with fits to Eq. \ref{LK}, are shown in Fig. \ref{PDO}(d). The fitted effective masses are $m_\alpha$ = 0.06(1) $m_e$, $m_\beta$ = 0.08(1) $m_e$, $m_\gamma$ = 0.26(2) $m_e$, $m_\delta$ = 1.2(1) $m_e$, and $m_\eta$ = 0.85(8) $m_e$. These values are comparable to both experimental masses in similar compounds \cite{EuAg4Sb2, CaCu4As2,SrAg4As2} and our calculations. 
\begin{table}
	\centering
	\caption {The parameters obtained from the quantum oscillation.}
	\label{Table_QS}
	\vskip .2cm
	\addtolength{\tabcolsep}{+2pt}
	\begin{tabular}{c c c c c c }
		\hline
		\hline
		& $F_\alpha$  & $F_\beta$ & $F_\gamma$ &$F_\delta$ &$F_\eta$ \\[0.5ex]
		\hline
		
		Frequency (T)              & 55           & 225           & 478           & 1410          & 1914   \\[1ex]
		$m^*$ ($m_e$)              & 0.06(1)      & 0.08(1)       & 0.26(2)       & 1.2(1)        & 0.85(8)   \\[1ex]
		$A_F$ (nm$^{-2}$)          & 0.52         & 2.15          & 4.56          & 13.45         & 18.26  \\[1ex]
		$k_F$ (nm$^{-1}$)          & 0.41         & 0.83          & 1.20          & 2.07          & 2.41   \\[1ex]
		$v_F$ (10$^6$ms$^{-1}$)    & 0.79         & 1.20          & 0.54          & 0.20          & 0.33   \\[1ex]
		
		\hline
		\hline
	\end{tabular}
\end{table}
\begin{figure*}
	\includegraphics[width=16.5cm, keepaspectratio]{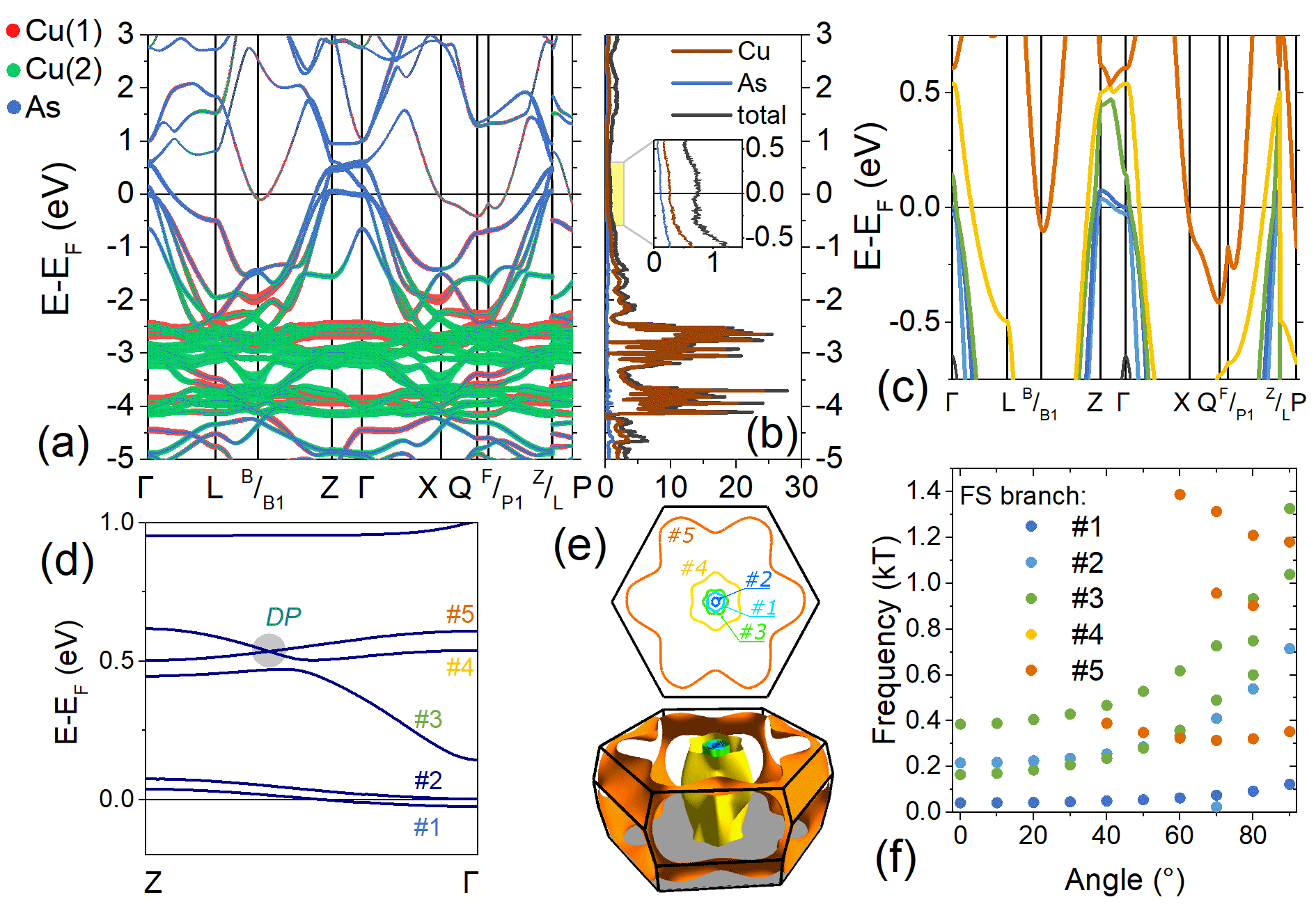}
	\caption{\label{BS-DOS} 
	(a) Band structure of \ce{SrCu4As2} with the contribution of the two inequivalent Cu atoms (Cu(1) \& Cu(2)) and As marked by the thickness of the red, green, and blue lines, respectively. Panel (b) shows the electronic density of states (DOS). Inset shows a close-up of the DOS around the Fermi energy, $E_F$ . Panel (c) shows the band dispersion around the Fermi energy. The 5 bands forming the Fermi surface are highlighted. Panel (d) shows the dispersion of the 5 bands along the $\Gamma$-$Z$ line, with band indices labeled using colors consistent with panel (c) and the Fermi-surface section coloring in panel (e). Panel (e) shows the Fermi surface of \ce{SrCu4As2} and its cross-section along the hexagonal [0001] direction[(the section plane is shown in gray]. Panel (f) shows theoretical SdH frequencies plotted against the angle between the hexagonal $c$ axis (parallel to the rhombohedral [111] direction) and $a$.}
\end{figure*}

The effective masses corresponding to bands $\alpha$ and $\beta$ are very small, suggesting that these Fermi pockets in \ce{SrCu4As2} have Dirac-like band dispersions. Bands with such low effective masses may play a role in causing the linear magnetoresistance seen in Fig. \ref{RT}. Measured effective masses, Fermi-surface cross-sectional areas ($A_F$) perpendicular to the applied magnetic field, Fermi wavevectors ($k_F$), and Fermi velocities ($v_F$) for each Fermi-surface section are summarized in Table \ref{Table_QS}. The Onsager relation $F = (\hbar/2\pi e)A_F$ is employed to determine the cross-sectional areas. Fermi wavevectors and velocities are calculated (assuming circular Fermi-surface cross-sections)  using the expressions $k_F = \sqrt{2eF/\hbar}$ and $v_F = \hbar k_F/m^*$, respectively, where $\hbar$ is the reduced Planck's constant.

Next, sample-orientation-dependent PDO measurements were performed to investigate the Fermi-surface topology of \ce{SrCu4As2}. The sample is rotated with respect to the $c$-axis, as illustrated in Fig. \ref{PDO}(e), in which $\theta$ = 0$^{\circ}$ corresponds to $H\parallel c$. Fig. \ref{PDO}(e) displays the SdH oscillation data at 690 mK for various angles. As $\theta$ increases, the oscillation amplitude slowly decreases, and at 90$^{\circ}$, the quantum oscillations disappear; these phenomena are also seen in the angle-dependent FFTs in Fig. \ref{PDO}(f). The angle dependences of the oscillation frequencies are displayed in Fig. \ref{PDO}(g). The frequency $F_{\alpha}$ is nearly independent of $\theta$, indicating an almost spherical three-dimensional Fermi-surface section. In contrast, $F_{\beta}$ and $F_{\gamma}$ exhibit $1/{\cos{\theta}}$ behavior, suggesting that these Fermi-surface sheets are quasi-two-dimensional. Notably, the higher frequencies $F_\delta$ and $F_\eta$ disappear very quickly as $\theta$ moves away from 0$^\circ$. Even at small angles of ±2.50$^\circ$, as shown in Fig. \ref{PDO}(h), these frequencies are not seen, which suggests that the Fermi-surface sections corresponding to bands $\delta$ and $\eta$  are strongly anisotropic.

\subsection{Electronic structure}

\begin{table*}[ht]
\centering
\caption{Calculated quantum oscillation frequencies and effective masses along the $c$-axis}
\label{tab:freq_mass}
\addtolength{\tabcolsep}{+18pt}
\begin{tabular}{@{} l *{5}{c} @{}}
\toprule
\text{Bands} & \textbf{\#1} & \textbf{\#2} & \textbf{\#3} & \textbf{\#4} & \textbf{\#5} \\
\hline
Frequency (\si{kT}) & 0.0420(1) & 0.216(4) & 
\begin{tabular}[t]{@{}c@{}}0.166(1)\\0.385(3)\end{tabular} & 
\begin{tabular}[t]{@{}c@{}}1.8144(2)\\2.3689(4)\end{tabular} & 
\begin{tabular}[t]{@{}c@{}}2.7712(3)\\17.592(6)\\28.394(6)\end{tabular} \\

Effective mass ($m_e$) & 0.120(4) & 0.28(2) & 
\begin{tabular}[t]{@{}c@{}}0.284(4)\\0.46(7)\end{tabular} & 
\begin{tabular}[t]{@{}c@{}}0.616(1)\\0.782(5)\end{tabular} & 
\begin{tabular}[t]{@{}c@{}}0.951(3)\\2.97(1)\\3.63(2)\end{tabular} \\
\toprule
\end{tabular}
\end{table*}

Figure \ref{BS-DOS}(a-d) show the calculated band structure and density of states (DOS) of \ce{SrCu4As2}. Most of the Cu $d$ state contribution to the DOS is limited to the narrow energy range between -4 to -2.5 eV, consistent with the Cu$^{+1}$ charge (3\textit{d}$^{10}$ 4\textit{s}$^0$ configuration). The DOS at the Fermi energy $E_F$ has contributions from interacting Cu and As $p$ atomic orbitals. In total, 5 bands cross the Fermi energy (Fig. \ref{BS-DOS}(c)). Two of them (\#1 and \#2 in Figs. \ref{BS-DOS}(d-f)) form small elliptical pockets centred around the $Z$ point of the Brillouin zone (BZ). Bands \#3, \#4, and \#5 form a tubular structure perpendicular to $\Gamma$-$Z$. Notably, bands \#4 and \#5 cross at a Dirac point located along the $\Gamma$-$Z$ line approximately 0.5 eV above the Fermi energy [Fig. \ref{BS-DOS}(d)]. Bands \#4 and \#5 cross the Fermi energy along different high symmetry lines: $\Gamma$-$Z$ and $B_1$-$Z$. Moreover, the DOS near the Fermi energy is low, which is consistent with the observed Hall carrier concentration. The gradients of the various bands crossing the Fermi energy suggest that \ce{SrCu4As2} is a semimetal, with holes dominant, in agreement with the Hall data. 

Shubnikov-de Haas oscillation frequencies obtained from the calculated Fermi surface are plotted in Fig. \ref{BS-DOS}(f) and summarized in Table \ref{tab:freq_mass}. Comparison with the experimentally measured SdH frequencies reveals that $F_{\alpha}$ can likely be ascribed to the first branch of the calculated Fermi surface, which is a small ellipsoidal pocket centered at the $Z$ point of the Brillouin zone, while $F_{\beta}$ matches the frequencies found in branches \#2 and \#3. One of the extremal orbits of the third branch can also be ascribed to $F_{\gamma}$. While the $ab$ $initio$ calculated and experimental SdH oscillations match qualitatively, the detailed frequencies and their angle dependence (especially at high tilt angles) differ. A major factor contributing to these differences is the fact that the calculations are done using the room-temperature crystal structure, whilst \ce{SrCu4As2} undergoes a structural phase transition at $T$ = 59 K, which affects the Fermi surface, as evidenced by a sharp change observed in carrier concentration. Another possible issue affecting the experimental data might be a small shift in the Fermi energy due to defects \cite{Ortiz2019,Gornicka2021,Lygouras2025} such as Sr vacancies suggested by the EDS spectroscopy.

\section{Summary}

We have grown high-quality single crystals of \ce{SrCu4As2} by a high-temperature solution growth-method using elemental Bi as a flux. The temperature-dependent electrical resistivity reveals a phase transition at a temperature $T_P$ = 59 K, lower than that reported in ref. \cite{ACu4As2_2023}. The origin of this phase transition, which exhibits hysteresis in the resistivity, remains unclear but it is most likely associated with a structural phase transition. Hall resistivity measurements indicate that the electrical transport in \ce{SrCu4As2} is dominated by the holes. The derived temperature-dependent carrier concentration shows a maximum near 50 K, presumably associated with the observed phase transition. In the studied crystals, a linear and nonsaturating magnetoresistance MR(H) is observed, with a relatively large MR (2 K, 9 T) = 120\%. Quantum-oscillation data measured in magnetic fields of up to 60 T reveal several oscillation frequencies with low effective masses, suggesting the presence of Dirac-like band dispersion in \ce{SrCu4As2}, a hypothesis that is supported by the band structure calculations. Finally, the angle-dependent SdH oscillation data and \textit{ab initio} calculations show a complex Fermi surface with multiple pockets, some of which have a quasi-two-dimensional character. Holelike pockets dominate, in agreement with the Hall data. 

\section{ACKNOWLEDGMENTS}

The work at Gdansk University of Technology was supported by the National Science Centre (Poland), Grant No. DEC-2024/08/X/ST3/00338. A portion of this work was performed at the National High Magnetic Field Laboratory, which is supported by National Science Foundation Cooperative Agreement No. DMR-2128556, the US Department of Energy (DoE)  and the State of Florida. JB and JS acknowledge support from the DoE BES FWP “Science of 100 T”, which permitted development of some of the high-field techniques used in the paper.

\bibliography{main}

\begin{thebibliography}{47}%
\makeatletter
\providecommand \@ifxundefined [1]{%
 \@ifx{#1\undefined}
}%
\providecommand \@ifnum [1]{%
 \ifnum #1\expandafter \@firstoftwo
 \else \expandafter \@secondoftwo
 \fi
}%
\providecommand \@ifx [1]{%
 \ifx #1\expandafter \@firstoftwo
 \else \expandafter \@secondoftwo
 \fi
}%
\providecommand \natexlab [1]{#1}%
\providecommand \enquote  [1]{``#1''}%
\providecommand \bibnamefont  [1]{#1}%
\providecommand \bibfnamefont [1]{#1}%
\providecommand \citenamefont [1]{#1}%
\providecommand \href@noop [0]{\@secondoftwo}%
\providecommand \href [0]{\begingroup \@sanitize@url \@href}%
\providecommand \@href[1]{\@@startlink{#1}\@@href}%
\providecommand \@@href[1]{\endgroup#1\@@endlink}%
\providecommand \@sanitize@url [0]{\catcode `\\12\catcode `\$12\catcode
  `\&12\catcode `\#12\catcode `\^12\catcode `\_12\catcode `\%12\relax}%
\providecommand \@@startlink[1]{}%
\providecommand \@@endlink[0]{}%
\providecommand \url  [0]{\begingroup\@sanitize@url \@url }%
\providecommand \@url [1]{\endgroup\@href {#1}{\urlprefix }}%
\providecommand \urlprefix  [0]{URL }%
\providecommand \Eprint [0]{\href }%
\providecommand \doibase [0]{https://doi.org/}%
\providecommand \selectlanguage [0]{\@gobble}%
\providecommand \bibinfo  [0]{\@secondoftwo}%
\providecommand \bibfield  [0]{\@secondoftwo}%
\providecommand \translation [1]{[#1]}%
\providecommand \BibitemOpen [0]{}%
\providecommand \bibitemStop [0]{}%
\providecommand \bibitemNoStop [0]{.\EOS\space}%
\providecommand \EOS [0]{\spacefactor3000\relax}%
\providecommand \BibitemShut  [1]{\csname bibitem#1\endcsname}%
\let\auto@bib@innerbib\@empty
%</preamble>
\bibitem [{\citenamefont {Du}\ \emph {et~al.}(2020)\citenamefont {Du},
  \citenamefont {Su}, \citenamefont {Luo}, \citenamefont {Shen}, \citenamefont
  {Nie}, \citenamefont {Yin}, \citenamefont {Chen}, \citenamefont {Li},
  \citenamefont {Smidman},\ and\ \citenamefont {Yuan}}]{LaAuSb2_2020}%
  \BibitemOpen
  \bibfield  {author} {\bibinfo {author} {\bibfnamefont {F.}~\bibnamefont
  {Du}}, \bibinfo {author} {\bibfnamefont {H.}~\bibnamefont {Su}}, \bibinfo
  {author} {\bibfnamefont {S.~S.}\ \bibnamefont {Luo}}, \bibinfo {author}
  {\bibfnamefont {B.}~\bibnamefont {Shen}}, \bibinfo {author} {\bibfnamefont
  {Z.~Y.}\ \bibnamefont {Nie}}, \bibinfo {author} {\bibfnamefont {L.~C.}\
  \bibnamefont {Yin}}, \bibinfo {author} {\bibfnamefont {Y.}~\bibnamefont
  {Chen}}, \bibinfo {author} {\bibfnamefont {R.}~\bibnamefont {Li}}, \bibinfo
  {author} {\bibfnamefont {M.}~\bibnamefont {Smidman}},\ and\ \bibinfo {author}
  {\bibfnamefont {H.~Q.}\ \bibnamefont {Yuan}},\ }\bibfield  {title} {\bibinfo
  {title} {{Interplay between charge density wave order and superconductivity
  in $\mathrm{La}\mathrm{Au}{\mathrm{Sb}}_{2}$ under pressure}},\ }\href
  {https://doi.org/10.1103/PhysRevB.102.144510} {\bibfield  {journal} {\bibinfo
   {journal} {Phys. Rev. B}\ }\textbf {\bibinfo {volume} {102}},\ \bibinfo
  {pages} {144510} (\bibinfo {year} {2020})}\BibitemShut {NoStop}%
\bibitem [{\citenamefont {Wu}\ \emph {et~al.}(2023)\citenamefont {Wu},
  \citenamefont {Hu}, \citenamefont {Graf}, \citenamefont {Liu}, \citenamefont
  {Deng}, \citenamefont {Fu}, \citenamefont {Kundu}, \citenamefont {Valla},
  \citenamefont {Petrovic},\ and\ \citenamefont {Wang}}]{LaAuSb2_2023}%
  \BibitemOpen
  \bibfield  {author} {\bibinfo {author} {\bibfnamefont {X.}~\bibnamefont
  {Wu}}, \bibinfo {author} {\bibfnamefont {Z.}~\bibnamefont {Hu}}, \bibinfo
  {author} {\bibfnamefont {D.}~\bibnamefont {Graf}}, \bibinfo {author}
  {\bibfnamefont {Y.}~\bibnamefont {Liu}}, \bibinfo {author} {\bibfnamefont
  {C.}~\bibnamefont {Deng}}, \bibinfo {author} {\bibfnamefont {H.}~\bibnamefont
  {Fu}}, \bibinfo {author} {\bibfnamefont {A.~K.}\ \bibnamefont {Kundu}},
  \bibinfo {author} {\bibfnamefont {T.}~\bibnamefont {Valla}}, \bibinfo
  {author} {\bibfnamefont {C.}~\bibnamefont {Petrovic}},\ and\ \bibinfo
  {author} {\bibfnamefont {A.}~\bibnamefont {Wang}},\ }\bibfield  {title}
  {\bibinfo {title} {{Coexistence of Dirac fermion and charge density wave in
  the square-net-based semimetal ${\mathrm{LaAuSb}}_{2}$}},\ }\href
  {https://doi.org/10.1103/PhysRevB.108.245156} {\bibfield  {journal} {\bibinfo
   {journal} {Phys. Rev. B}\ }\textbf {\bibinfo {volume} {108}},\ \bibinfo
  {pages} {245156} (\bibinfo {year} {2023})}\BibitemShut {NoStop}%
\bibitem [{\citenamefont {Shi}\ \emph {et~al.}(2016)\citenamefont {Shi},
  \citenamefont {Richard}, \citenamefont {Wang}, \citenamefont {Liu},
  \citenamefont {Matt}, \citenamefont {Xu}, \citenamefont {Dhaka},
  \citenamefont {Ristic}, \citenamefont {Qian}, \citenamefont {Yang},
  \citenamefont {Petrovic}, \citenamefont {Shi},\ and\ \citenamefont
  {Ding}}]{LaAgSb2_2016}%
  \BibitemOpen
  \bibfield  {author} {\bibinfo {author} {\bibfnamefont {X.}~\bibnamefont
  {Shi}}, \bibinfo {author} {\bibfnamefont {P.}~\bibnamefont {Richard}},
  \bibinfo {author} {\bibfnamefont {K.}~\bibnamefont {Wang}}, \bibinfo {author}
  {\bibfnamefont {M.}~\bibnamefont {Liu}}, \bibinfo {author} {\bibfnamefont
  {C.~E.}\ \bibnamefont {Matt}}, \bibinfo {author} {\bibfnamefont
  {N.}~\bibnamefont {Xu}}, \bibinfo {author} {\bibfnamefont {R.~S.}\
  \bibnamefont {Dhaka}}, \bibinfo {author} {\bibfnamefont {Z.}~\bibnamefont
  {Ristic}}, \bibinfo {author} {\bibfnamefont {T.}~\bibnamefont {Qian}},
  \bibinfo {author} {\bibfnamefont {Y.-F.}\ \bibnamefont {Yang}}, \bibinfo
  {author} {\bibfnamefont {C.}~\bibnamefont {Petrovic}}, \bibinfo {author}
  {\bibfnamefont {M.}~\bibnamefont {Shi}},\ and\ \bibinfo {author}
  {\bibfnamefont {H.}~\bibnamefont {Ding}},\ }\bibfield  {title} {\bibinfo
  {title} {{Observation of Dirac-like band dispersion in
  ${\mathrm{LaAgSb}}_{2}$}},\ }\href
  {https://doi.org/10.1103/PhysRevB.93.081105} {\bibfield  {journal} {\bibinfo
  {journal} {Phys. Rev. B}\ }\textbf {\bibinfo {volume} {93}},\ \bibinfo
  {pages} {081105} (\bibinfo {year} {2016})}\BibitemShut {NoStop}%
\bibitem [{\citenamefont {Bud'ko}\ \emph {et~al.}(2006)\citenamefont {Bud'ko},
  \citenamefont {Wiener}, \citenamefont {Ribeiro}, \citenamefont {Canfield},
  \citenamefont {Lee}, \citenamefont {Vogt},\ and\ \citenamefont
  {Lacerda}}]{LaAgSb2}%
  \BibitemOpen
  \bibfield  {author} {\bibinfo {author} {\bibfnamefont {S.~L.}\ \bibnamefont
  {Bud'ko}}, \bibinfo {author} {\bibfnamefont {T.~A.}\ \bibnamefont {Wiener}},
  \bibinfo {author} {\bibfnamefont {R.~A.}\ \bibnamefont {Ribeiro}}, \bibinfo
  {author} {\bibfnamefont {P.~C.}\ \bibnamefont {Canfield}}, \bibinfo {author}
  {\bibfnamefont {Y.}~\bibnamefont {Lee}}, \bibinfo {author} {\bibfnamefont
  {T.}~\bibnamefont {Vogt}},\ and\ \bibinfo {author} {\bibfnamefont {A.~H.}\
  \bibnamefont {Lacerda}},\ }\bibfield  {title} {\bibinfo {title} {{Effect of
  pressure and chemical substitutions on the charge-density-wave in
  ${\mathrm{LaAgSb}}_{2}$}},\ }\href
  {https://doi.org/10.1103/PhysRevB.73.184111} {\bibfield  {journal} {\bibinfo
  {journal} {Phys. Rev. B}\ }\textbf {\bibinfo {volume} {73}},\ \bibinfo
  {pages} {184111} (\bibinfo {year} {2006})}\BibitemShut {NoStop}%
\bibitem [{\citenamefont {Malick}\ \emph {et~al.}(2025)\citenamefont {Malick},
  \citenamefont {\ifmmode~\acute{S}\else \'{S}\fi{}wi\k{a}tek}, \citenamefont
  {Winiarski},\ and\ \citenamefont {Klimczuk}}]{PrAgBi2}%
  \BibitemOpen
  \bibfield  {author} {\bibinfo {author} {\bibfnamefont {S.}~\bibnamefont
  {Malick}}, \bibinfo {author} {\bibfnamefont {H.}~\bibnamefont
  {\ifmmode~\acute{S}\else \'{S}\fi{}wi\k{a}tek}}, \bibinfo {author}
  {\bibfnamefont {M.~J.}\ \bibnamefont {Winiarski}},\ and\ \bibinfo {author}
  {\bibfnamefont {T.}~\bibnamefont {Klimczuk}},\ }\bibfield  {title} {\bibinfo
  {title} {{Observation of quantum oscillations, linear magnetoresistance, and
  crystalline electric field effect in quasi-two-dimensional
  ${\mathrm{PrAgBi}}_{2}$}},\ }\href
  {https://doi.org/10.1103/PhysRevB.111.045144} {\bibfield  {journal} {\bibinfo
   {journal} {Phys. Rev. B}\ }\textbf {\bibinfo {volume} {111}},\ \bibinfo
  {pages} {045144} (\bibinfo {year} {2025})}\BibitemShut {NoStop}%
\bibitem [{\citenamefont {Lou}\ \emph {et~al.}(2020)\citenamefont {Lou},
  \citenamefont {Xu}, \citenamefont {Wen}, \citenamefont {Yu}, \citenamefont
  {Wei}, \citenamefont {Yao}, \citenamefont {Song}, \citenamefont
  {Emmanouilidou}, \citenamefont {Shen}, \citenamefont {Ni}, \citenamefont
  {Dudin}, \citenamefont {Huang}, \citenamefont {Denlinger}, \citenamefont
  {Sutarto}, \citenamefont {Li}, \citenamefont {Peng},\ and\ \citenamefont
  {Feng}}]{BaAg2As2}%
  \BibitemOpen
  \bibfield  {author} {\bibinfo {author} {\bibfnamefont {X.}~\bibnamefont
  {Lou}}, \bibinfo {author} {\bibfnamefont {H.~C.}\ \bibnamefont {Xu}},
  \bibinfo {author} {\bibfnamefont {C.~H.~P.}\ \bibnamefont {Wen}}, \bibinfo
  {author} {\bibfnamefont {T.~L.}\ \bibnamefont {Yu}}, \bibinfo {author}
  {\bibfnamefont {W.~Z.}\ \bibnamefont {Wei}}, \bibinfo {author} {\bibfnamefont
  {Q.}~\bibnamefont {Yao}}, \bibinfo {author} {\bibfnamefont {Y.~H.}\
  \bibnamefont {Song}}, \bibinfo {author} {\bibfnamefont {E.}~\bibnamefont
  {Emmanouilidou}}, \bibinfo {author} {\bibfnamefont {B.}~\bibnamefont {Shen}},
  \bibinfo {author} {\bibfnamefont {N.}~\bibnamefont {Ni}}, \bibinfo {author}
  {\bibfnamefont {P.}~\bibnamefont {Dudin}}, \bibinfo {author} {\bibfnamefont
  {Y.~B.}\ \bibnamefont {Huang}}, \bibinfo {author} {\bibfnamefont
  {J.}~\bibnamefont {Denlinger}}, \bibinfo {author} {\bibfnamefont
  {R.}~\bibnamefont {Sutarto}}, \bibinfo {author} {\bibfnamefont
  {W.}~\bibnamefont {Li}}, \bibinfo {author} {\bibfnamefont {R.}~\bibnamefont
  {Peng}},\ and\ \bibinfo {author} {\bibfnamefont {D.~L.}\ \bibnamefont
  {Feng}},\ }\bibfield  {title} {\bibinfo {title} {{Lattice distortion and
  electronic structure of ${\mathrm{BaAg}}_{2}{\mathrm{As}}_{2}$ across its
  nonmagnetic phase transition}},\ }\href
  {https://doi.org/10.1103/PhysRevB.101.075123} {\bibfield  {journal} {\bibinfo
   {journal} {Phys. Rev. B}\ }\textbf {\bibinfo {volume} {101}},\ \bibinfo
  {pages} {075123} (\bibinfo {year} {2020})}\BibitemShut {NoStop}%
\bibitem [{\citenamefont {Huang}\ \emph {et~al.}(2008)\citenamefont {Huang},
  \citenamefont {Qiu}, \citenamefont {Bao}, \citenamefont {Green},
  \citenamefont {Lynn}, \citenamefont {Gasparovic}, \citenamefont {Wu},
  \citenamefont {Wu},\ and\ \citenamefont {Chen}}]{BaFe2As2}%
  \BibitemOpen
  \bibfield  {author} {\bibinfo {author} {\bibfnamefont {Q.}~\bibnamefont
  {Huang}}, \bibinfo {author} {\bibfnamefont {Y.}~\bibnamefont {Qiu}}, \bibinfo
  {author} {\bibfnamefont {W.}~\bibnamefont {Bao}}, \bibinfo {author}
  {\bibfnamefont {M.~A.}\ \bibnamefont {Green}}, \bibinfo {author}
  {\bibfnamefont {J.~W.}\ \bibnamefont {Lynn}}, \bibinfo {author}
  {\bibfnamefont {Y.~C.}\ \bibnamefont {Gasparovic}}, \bibinfo {author}
  {\bibfnamefont {T.}~\bibnamefont {Wu}}, \bibinfo {author} {\bibfnamefont
  {G.}~\bibnamefont {Wu}},\ and\ \bibinfo {author} {\bibfnamefont {X.~H.}\
  \bibnamefont {Chen}},\ }\bibfield  {title} {\bibinfo {title}
  {{Neutron-Diffraction Measurements of Magnetic Order and a Structural
  Transition in the Parent ${\mathrm{BaFe}}_{2}{\mathrm{As}}_{2}$ Compound of
  FeAs-Based High-Temperature Superconductors}},\ }\href
  {https://doi.org/10.1103/PhysRevLett.101.257003} {\bibfield  {journal}
  {\bibinfo  {journal} {Phys. Rev. Lett.}\ }\textbf {\bibinfo {volume} {101}},\
  \bibinfo {pages} {257003} (\bibinfo {year} {2008})}\BibitemShut {NoStop}%
\bibitem [{\citenamefont {Huynh}\ \emph {et~al.}(2011)\citenamefont {Huynh},
  \citenamefont {Tanabe},\ and\ \citenamefont {Tanigaki}}]{BaFe2As2_2011}%
  \BibitemOpen
  \bibfield  {author} {\bibinfo {author} {\bibfnamefont {K.~K.}\ \bibnamefont
  {Huynh}}, \bibinfo {author} {\bibfnamefont {Y.}~\bibnamefont {Tanabe}},\ and\
  \bibinfo {author} {\bibfnamefont {K.}~\bibnamefont {Tanigaki}},\ }\bibfield
  {title} {\bibinfo {title} {{Both Electron and Hole Dirac Cone States in
  $\mathrm{Ba}(\mathrm{FeAs}{)}_{2}$ Confirmed by Magnetoresistance}},\ }\href
  {https://doi.org/10.1103/PhysRevLett.106.217004} {\bibfield  {journal}
  {\bibinfo  {journal} {Phys. Rev. Lett.}\ }\textbf {\bibinfo {volume} {106}},\
  \bibinfo {pages} {217004} (\bibinfo {year} {2011})}\BibitemShut {NoStop}%
\bibitem [{\citenamefont {Malick}\ \emph {et~al.}(2024)\citenamefont {Malick},
  \citenamefont {\ifmmode~\acute{S}\else \'{S}\fi{}wi\k{a}tek}, \citenamefont
  {B\l{}awat}, \citenamefont {Singleton},\ and\ \citenamefont
  {Klimczuk}}]{EuAg4Sb2}%
  \BibitemOpen
  \bibfield  {author} {\bibinfo {author} {\bibfnamefont {S.}~\bibnamefont
  {Malick}}, \bibinfo {author} {\bibfnamefont {H.}~\bibnamefont
  {\ifmmode~\acute{S}\else \'{S}\fi{}wi\k{a}tek}}, \bibinfo {author}
  {\bibfnamefont {J.}~\bibnamefont {B\l{}awat}}, \bibinfo {author}
  {\bibfnamefont {J.}~\bibnamefont {Singleton}},\ and\ \bibinfo {author}
  {\bibfnamefont {T.}~\bibnamefont {Klimczuk}},\ }\bibfield  {title} {\bibinfo
  {title} {{Large magnetoresistance and first-order phase transition in
  antiferromagnetic single-crystalline
  $\mathrm{EuAg}{}_{4}\mathrm{Sb}{}_{2}$}},\ }\href
  {https://doi.org/10.1103/PhysRevB.110.165149} {\bibfield  {journal} {\bibinfo
   {journal} {Phys. Rev. B}\ }\textbf {\bibinfo {volume} {110}},\ \bibinfo
  {pages} {165149} (\bibinfo {year} {2024})}\BibitemShut {NoStop}%
\bibitem [{\citenamefont {Sasmal}\ \emph {et~al.}(2022)\citenamefont {Sasmal},
  \citenamefont {Saini}, \citenamefont {Ramakrishnan}, \citenamefont {Dwari},
  \citenamefont {Maity}, \citenamefont {Bao}, \citenamefont {Mondal},
  \citenamefont {Tripathi}, \citenamefont {van Smaalen}, \citenamefont
  {Singh},\ and\ \citenamefont {Thamizhavel}}]{CaCu4As2}%
  \BibitemOpen
  \bibfield  {author} {\bibinfo {author} {\bibfnamefont {S.}~\bibnamefont
  {Sasmal}}, \bibinfo {author} {\bibfnamefont {V.}~\bibnamefont {Saini}},
  \bibinfo {author} {\bibfnamefont {S.}~\bibnamefont {Ramakrishnan}}, \bibinfo
  {author} {\bibfnamefont {G.}~\bibnamefont {Dwari}}, \bibinfo {author}
  {\bibfnamefont {B.~B.}\ \bibnamefont {Maity}}, \bibinfo {author}
  {\bibfnamefont {J.-K.}\ \bibnamefont {Bao}}, \bibinfo {author} {\bibfnamefont
  {R.}~\bibnamefont {Mondal}}, \bibinfo {author} {\bibfnamefont
  {V.}~\bibnamefont {Tripathi}}, \bibinfo {author} {\bibfnamefont
  {S.}~\bibnamefont {van Smaalen}}, \bibinfo {author} {\bibfnamefont
  {B.}~\bibnamefont {Singh}},\ and\ \bibinfo {author} {\bibfnamefont
  {A.}~\bibnamefont {Thamizhavel}},\ }\bibfield  {title} {\bibinfo {title}
  {{Observation of multilayer quantum Hall effect in the charge density wave
  material ${\mathrm{CaCu}}_{4}{\mathrm{As}}_{2}$}},\ }\href
  {https://doi.org/10.1103/PhysRevResearch.4.L012011} {\bibfield  {journal}
  {\bibinfo  {journal} {Phys. Rev. Res.}\ }\textbf {\bibinfo {volume} {4}},\
  \bibinfo {pages} {L012011} (\bibinfo {year} {2022})}\BibitemShut {NoStop}%
\bibitem [{\citenamefont {Kurumaji}\ \emph {et~al.}(2025)\citenamefont
  {Kurumaji}, \citenamefont {Paul}, \citenamefont {Fang}, \citenamefont
  {Neves}, \citenamefont {Kang}, \citenamefont {White}, \citenamefont
  {Nakajima}, \citenamefont {Graf}, \citenamefont {Ye}, \citenamefont {Chan},
  \citenamefont {Suzuki}, \citenamefont {Denlinger}, \citenamefont {Jozwiak},
  \citenamefont {Bostwick}, \citenamefont {Rotenberg}, \citenamefont {Zhao},
  \citenamefont {Lynn}, \citenamefont {Kaxiras}, \citenamefont {Comin},
  \citenamefont {Fu},\ and\ \citenamefont {Checkelsky}}]{EuAg4Sb2_sciadv}%
  \BibitemOpen
  \bibfield  {author} {\bibinfo {author} {\bibfnamefont {T.}~\bibnamefont
  {Kurumaji}}, \bibinfo {author} {\bibfnamefont {N.}~\bibnamefont {Paul}},
  \bibinfo {author} {\bibfnamefont {S.}~\bibnamefont {Fang}}, \bibinfo {author}
  {\bibfnamefont {P.~M.}\ \bibnamefont {Neves}}, \bibinfo {author}
  {\bibfnamefont {M.}~\bibnamefont {Kang}}, \bibinfo {author} {\bibfnamefont
  {J.~S.}\ \bibnamefont {White}}, \bibinfo {author} {\bibfnamefont
  {T.}~\bibnamefont {Nakajima}}, \bibinfo {author} {\bibfnamefont
  {D.}~\bibnamefont {Graf}}, \bibinfo {author} {\bibfnamefont {L.}~\bibnamefont
  {Ye}}, \bibinfo {author} {\bibfnamefont {M.~K.}\ \bibnamefont {Chan}},
  \bibinfo {author} {\bibfnamefont {T.}~\bibnamefont {Suzuki}}, \bibinfo
  {author} {\bibfnamefont {J.}~\bibnamefont {Denlinger}}, \bibinfo {author}
  {\bibfnamefont {C.}~\bibnamefont {Jozwiak}}, \bibinfo {author} {\bibfnamefont
  {A.}~\bibnamefont {Bostwick}}, \bibinfo {author} {\bibfnamefont
  {E.}~\bibnamefont {Rotenberg}}, \bibinfo {author} {\bibfnamefont
  {Y.}~\bibnamefont {Zhao}}, \bibinfo {author} {\bibfnamefont {J.~W.}\
  \bibnamefont {Lynn}}, \bibinfo {author} {\bibfnamefont {E.}~\bibnamefont
  {Kaxiras}}, \bibinfo {author} {\bibfnamefont {R.}~\bibnamefont {Comin}},
  \bibinfo {author} {\bibfnamefont {L.}~\bibnamefont {Fu}},\ and\ \bibinfo
  {author} {\bibfnamefont {J.~G.}\ \bibnamefont {Checkelsky}},\ }\bibfield
  {title} {\bibinfo {title} {{Electronic commensuration of a spin moiré
  superlattice in a layered magnetic semimetal}},\ }\href
  {https://doi.org/10.1126/sciadv.adu6686} {\bibfield  {journal} {\bibinfo
  {journal} {Science Advances}\ }\textbf {\bibinfo {volume} {11}},\ \bibinfo
  {pages} {eadu6686} (\bibinfo {year} {2025})}\BibitemShut {NoStop}%
\bibitem [{\citenamefont {Shen}\ \emph {et~al.}(2020)\citenamefont {Shen},
  \citenamefont {Hu}, \citenamefont {Cao}, \citenamefont {Gui}, \citenamefont
  {Emmanouilidou}, \citenamefont {Xie},\ and\ \citenamefont
  {Ni}}]{EuAg4As2_2020}%
  \BibitemOpen
  \bibfield  {author} {\bibinfo {author} {\bibfnamefont {B.}~\bibnamefont
  {Shen}}, \bibinfo {author} {\bibfnamefont {C.}~\bibnamefont {Hu}}, \bibinfo
  {author} {\bibfnamefont {H.}~\bibnamefont {Cao}}, \bibinfo {author}
  {\bibfnamefont {X.}~\bibnamefont {Gui}}, \bibinfo {author} {\bibfnamefont
  {E.}~\bibnamefont {Emmanouilidou}}, \bibinfo {author} {\bibfnamefont
  {W.}~\bibnamefont {Xie}},\ and\ \bibinfo {author} {\bibfnamefont
  {N.}~\bibnamefont {Ni}},\ }\bibfield  {title} {\bibinfo {title} {{Structural
  distortion and incommensurate noncollinear magnetism in
  ${\mathrm{EuAg}}_{4}{\mathrm{As}}_{2}$}},\ }\href
  {https://doi.org/10.1103/PhysRevMaterials.4.064419} {\bibfield  {journal}
  {\bibinfo  {journal} {Phys. Rev. Mater.}\ }\textbf {\bibinfo {volume} {4}},\
  \bibinfo {pages} {064419} (\bibinfo {year} {2020})}\BibitemShut {NoStop}%
\bibitem [{\citenamefont {Zhu}\ \emph {et~al.}(2021)\citenamefont {Zhu},
  \citenamefont {Li}, \citenamefont {Yang}, \citenamefont {Lou}, \citenamefont
  {Du}, \citenamefont {Yang}, \citenamefont {Chen}, \citenamefont {Wang},\ and\
  \citenamefont {Fang}}]{EuAg4As2_2021}%
  \BibitemOpen
  \bibfield  {author} {\bibinfo {author} {\bibfnamefont {Q.}~\bibnamefont
  {Zhu}}, \bibinfo {author} {\bibfnamefont {L.}~\bibnamefont {Li}}, \bibinfo
  {author} {\bibfnamefont {Z.}~\bibnamefont {Yang}}, \bibinfo {author}
  {\bibfnamefont {Z.}~\bibnamefont {Lou}}, \bibinfo {author} {\bibfnamefont
  {J.}~\bibnamefont {Du}}, \bibinfo {author} {\bibfnamefont {J.}~\bibnamefont
  {Yang}}, \bibinfo {author} {\bibfnamefont {B.}~\bibnamefont {Chen}}, \bibinfo
  {author} {\bibfnamefont {H.}~\bibnamefont {Wang}},\ and\ \bibinfo {author}
  {\bibfnamefont {M.}~\bibnamefont {Fang}},\ }\bibfield  {title} {\bibinfo
  {title} {{Metamagnetic transitions and anomalous magnetoresistance in
  EuAg$_4$As$_2$ crystals}},\ }\href
  {https://doi.org/https://doi.org/10.1007/s11433-020-1629-x} {\bibfield
  {journal} {\bibinfo  {journal} {Science China Physics, Mechanics \&
  Astronomy}\ }\textbf {\bibinfo {volume} {64}},\ \bibinfo {pages} {227011}
  (\bibinfo {year} {2021})}\BibitemShut {NoStop}%
\bibitem [{\citenamefont {Li}\ \emph {et~al.}(2022)\citenamefont {Li},
  \citenamefont {Yang}, \citenamefont {Su}, \citenamefont {Yang}, \citenamefont
  {Chen}, \citenamefont {Du}, \citenamefont {Wu}, \citenamefont {Wang},\ and\
  \citenamefont {Fang}}]{EuCu4As2}%
  \BibitemOpen
  \bibfield  {author} {\bibinfo {author} {\bibfnamefont {L.}~\bibnamefont
  {Li}}, \bibinfo {author} {\bibfnamefont {Z.}~\bibnamefont {Yang}}, \bibinfo
  {author} {\bibfnamefont {Q.}~\bibnamefont {Su}}, \bibinfo {author}
  {\bibfnamefont {J.}~\bibnamefont {Yang}}, \bibinfo {author} {\bibfnamefont
  {B.}~\bibnamefont {Chen}}, \bibinfo {author} {\bibfnamefont {J.}~\bibnamefont
  {Du}}, \bibinfo {author} {\bibfnamefont {C.}~\bibnamefont {Wu}}, \bibinfo
  {author} {\bibfnamefont {H.}~\bibnamefont {Wang}},\ and\ \bibinfo {author}
  {\bibfnamefont {M.}~\bibnamefont {Fang}},\ }\bibfield  {title} {\bibinfo
  {title} {{Large positive and negative magnetoresistance in the magnetic
  EuCu$_4$As$_2$ crystal}},\ }\href
  {https://doi.org/https://doi.org/10.1016/j.jallcom.2022.165460} {\bibfield
  {journal} {\bibinfo  {journal} {Journal of Alloys and Compounds}\ }\textbf
  {\bibinfo {volume} {916}},\ \bibinfo {pages} {165460} (\bibinfo {year}
  {2022})}\BibitemShut {NoStop}%
\bibitem [{\citenamefont {Shen}\ \emph {et~al.}(2018)\citenamefont {Shen},
  \citenamefont {Emmanouilidou}, \citenamefont {Deng}, \citenamefont
  {McCollam}, \citenamefont {Xing}, \citenamefont {Kotliar}, \citenamefont
  {Coldea},\ and\ \citenamefont {Ni}}]{SrAg4As2}%
  \BibitemOpen
  \bibfield  {author} {\bibinfo {author} {\bibfnamefont {B.}~\bibnamefont
  {Shen}}, \bibinfo {author} {\bibfnamefont {E.}~\bibnamefont {Emmanouilidou}},
  \bibinfo {author} {\bibfnamefont {X.}~\bibnamefont {Deng}}, \bibinfo {author}
  {\bibfnamefont {A.}~\bibnamefont {McCollam}}, \bibinfo {author}
  {\bibfnamefont {J.}~\bibnamefont {Xing}}, \bibinfo {author} {\bibfnamefont
  {G.}~\bibnamefont {Kotliar}}, \bibinfo {author} {\bibfnamefont {A.~I.}\
  \bibnamefont {Coldea}},\ and\ \bibinfo {author} {\bibfnamefont
  {N.}~\bibnamefont {Ni}},\ }\bibfield  {title} {\bibinfo {title} {{Significant
  change in the electronic behavior associated with structural distortions in
  monocrystalline ${\mathrm{SrAg}}_{4}{\mathrm{As}}_{2}$}},\ }\href
  {https://doi.org/10.1103/PhysRevB.98.235130} {\bibfield  {journal} {\bibinfo
  {journal} {Phys. Rev. B}\ }\textbf {\bibinfo {volume} {98}},\ \bibinfo
  {pages} {235130} (\bibinfo {year} {2018})}\BibitemShut {NoStop}%
\bibitem [{\citenamefont {Dünner}\ and\ \citenamefont
  {Mewis}(1999)}]{SrCu4As2_1999}%
  \BibitemOpen
  \bibfield  {author} {\bibinfo {author} {\bibfnamefont {J.}~\bibnamefont
  {Dünner}}\ and\ \bibinfo {author} {\bibfnamefont {A.}~\bibnamefont
  {Mewis}},\ }\bibfield  {title} {\bibinfo {title} {{Synthese und
  Kristallstruktur von ACu$_4$As$_2$ (A: Ca–Ba, Eu)}},\ }\href
  {https://doi.org/https://doi.org/10.1002/(SICI)1521-3749(199904)625:4<625::AID-ZAAC625>3.0.CO;2-Z}
  {\bibfield  {journal} {\bibinfo  {journal} {Zeitschrift für anorganische und
  allgemeine Chemie}\ }\textbf {\bibinfo {volume} {625}},\ \bibinfo {pages}
  {625} (\bibinfo {year} {1999})}\BibitemShut {NoStop}%
\bibitem [{\citenamefont {Nie}\ \emph {et~al.}(2023)\citenamefont {Nie},
  \citenamefont {Chen}, \citenamefont {Mei}, \citenamefont {Wang},
  \citenamefont {Wu}, \citenamefont {Jiang}, \citenamefont {Song},
  \citenamefont {Ning}, \citenamefont {Wang}, \citenamefont {Zhu},\ and\
  \citenamefont {Tian}}]{ACu4As2_2023}%
  \BibitemOpen
  \bibfield  {author} {\bibinfo {author} {\bibfnamefont {Y.}~\bibnamefont
  {Nie}}, \bibinfo {author} {\bibfnamefont {Z.}~\bibnamefont {Chen}}, \bibinfo
  {author} {\bibfnamefont {M.}~\bibnamefont {Mei}}, \bibinfo {author}
  {\bibfnamefont {Y.-Y.}\ \bibnamefont {Wang}}, \bibinfo {author}
  {\bibfnamefont {J.-T.}\ \bibnamefont {Wu}}, \bibinfo {author} {\bibfnamefont
  {J.-L.}\ \bibnamefont {Jiang}}, \bibinfo {author} {\bibfnamefont {W.-H.}\
  \bibnamefont {Song}}, \bibinfo {author} {\bibfnamefont {W.}~\bibnamefont
  {Ning}}, \bibinfo {author} {\bibfnamefont {Z.-S.}\ \bibnamefont {Wang}},
  \bibinfo {author} {\bibfnamefont {X.-D.}\ \bibnamefont {Zhu}},\ and\ \bibinfo
  {author} {\bibfnamefont {M.-L.}\ \bibnamefont {Tian}},\ }\bibfield  {title}
  {\bibinfo {title} {{Subtle lattice distortion-driven phase transitions in
  layered ACu$_4$As$_2$ (A = Eu, Sr)}},\ }\href
  {https://doi.org/10.1088/1674-1056/acd36b} {\bibfield  {journal} {\bibinfo
  {journal} {Chinese Physics B}\ }\textbf {\bibinfo {volume} {32}},\ \bibinfo
  {pages} {106102} (\bibinfo {year} {2023})}\BibitemShut {NoStop}%
\bibitem [{\citenamefont {Hadjiev}\ \emph {et~al.}(2011)\citenamefont
  {Hadjiev}, \citenamefont {Lv},\ and\ \citenamefont {Chu}}]{SrCu4AS2_DFT}%
  \BibitemOpen
  \bibfield  {author} {\bibinfo {author} {\bibfnamefont {V.~G.}\ \bibnamefont
  {Hadjiev}}, \bibinfo {author} {\bibfnamefont {B.}~\bibnamefont {Lv}},\ and\
  \bibinfo {author} {\bibfnamefont {C.~W.}\ \bibnamefont {Chu}},\ }\bibfield
  {title} {\bibinfo {title} {{Electronic band structure of
  SrCu${}_{4}$As${}_{2}$ and KCu${}_{4}$As${}_{2}$: Metals with diversely doped
  CuAs layers}},\ }\href {https://doi.org/10.1103/PhysRevB.84.073105}
  {\bibfield  {journal} {\bibinfo  {journal} {Phys. Rev. B}\ }\textbf {\bibinfo
  {volume} {84}},\ \bibinfo {pages} {073105} (\bibinfo {year}
  {2011})}\BibitemShut {NoStop}%
\bibitem [{SM()}]{SM}%
  \BibitemOpen
  \href@noop {} {\bibinfo {title} {{See Supplemental Material for detailed
  single crystal growth, structural chracterization, heat capacity data, FFT of
  PDO data, band structure calculations}}}\BibitemShut {NoStop}%
\bibitem [{\citenamefont {Ghannadzadeh}\ \emph {et~al.}(2011)\citenamefont
  {Ghannadzadeh}, \citenamefont {Coak}, \citenamefont {Franke}, \citenamefont
  {Goddard}, \citenamefont {Singleton},\ and\ \citenamefont
  {Manson}}]{PDO_Exp}%
  \BibitemOpen
  \bibfield  {author} {\bibinfo {author} {\bibfnamefont {S.}~\bibnamefont
  {Ghannadzadeh}}, \bibinfo {author} {\bibfnamefont {M.}~\bibnamefont {Coak}},
  \bibinfo {author} {\bibfnamefont {I.}~\bibnamefont {Franke}}, \bibinfo
  {author} {\bibfnamefont {P.~A.}\ \bibnamefont {Goddard}}, \bibinfo {author}
  {\bibfnamefont {J.}~\bibnamefont {Singleton}},\ and\ \bibinfo {author}
  {\bibfnamefont {J.~L.}\ \bibnamefont {Manson}},\ }\bibfield  {title}
  {\bibinfo {title} {{Measurement of magnetic susceptibility in pulsed magnetic
  fields using a proximity detector oscillator}},\ }\href
  {https://doi.org/10.1063/1.3653395} {\bibfield  {journal} {\bibinfo
  {journal} {Review of Scientific Instruments}\ }\textbf {\bibinfo {volume}
  {82}},\ \bibinfo {pages} {113902} (\bibinfo {year} {2011})}\BibitemShut
  {NoStop}%
\bibitem [{\citenamefont {G\"otze}\ \emph {et~al.}(2020)\citenamefont
  {G\"otze}, \citenamefont {Pearce}, \citenamefont {Goddard}, \citenamefont
  {Jaime}, \citenamefont {Maple}, \citenamefont {Sasmal}, \citenamefont
  {Yanagisawa}, \citenamefont {McCollam}, \citenamefont {Khouri}, \citenamefont
  {Ho},\ and\ \citenamefont {Singleton}}]{CeOs4Sb12}%
  \BibitemOpen
  \bibfield  {author} {\bibinfo {author} {\bibfnamefont {K.}~\bibnamefont
  {G\"otze}}, \bibinfo {author} {\bibfnamefont {M.~J.}\ \bibnamefont {Pearce}},
  \bibinfo {author} {\bibfnamefont {P.~A.}\ \bibnamefont {Goddard}}, \bibinfo
  {author} {\bibfnamefont {M.}~\bibnamefont {Jaime}}, \bibinfo {author}
  {\bibfnamefont {M.~B.}\ \bibnamefont {Maple}}, \bibinfo {author}
  {\bibfnamefont {K.}~\bibnamefont {Sasmal}}, \bibinfo {author} {\bibfnamefont
  {T.}~\bibnamefont {Yanagisawa}}, \bibinfo {author} {\bibfnamefont
  {A.}~\bibnamefont {McCollam}}, \bibinfo {author} {\bibfnamefont
  {T.}~\bibnamefont {Khouri}}, \bibinfo {author} {\bibfnamefont {P.-C.}\
  \bibnamefont {Ho}},\ and\ \bibinfo {author} {\bibfnamefont {J.}~\bibnamefont
  {Singleton}},\ }\bibfield  {title} {\bibinfo {title} {{Unusual phase boundary
  of the magnetic-field-tuned valence transition in
  ${\mathrm{CeOs}}_{4}{\mathrm{Sb}}_{12}$}},\ }\href
  {https://doi.org/10.1103/PhysRevB.101.075102} {\bibfield  {journal} {\bibinfo
   {journal} {Phys. Rev. B}\ }\textbf {\bibinfo {volume} {101}},\ \bibinfo
  {pages} {075102} (\bibinfo {year} {2020})}\BibitemShut {NoStop}%
\bibitem [{elk()}]{elk}%
  \BibitemOpen
  \href@noop {} {\bibinfo {title} {{The Elk Code}}},\ \bibinfo {howpublished}
  {\url{http://elk.sourceforge.net/}}\BibitemShut {NoStop}%
\bibitem [{\citenamefont {Perdew}\ \emph {et~al.}(1996)\citenamefont {Perdew},
  \citenamefont {Burke},\ and\ \citenamefont {Ernzerhof}}]{Model_77}%
  \BibitemOpen
  \bibfield  {author} {\bibinfo {author} {\bibfnamefont {J.~P.}\ \bibnamefont
  {Perdew}}, \bibinfo {author} {\bibfnamefont {K.}~\bibnamefont {Burke}},\ and\
  \bibinfo {author} {\bibfnamefont {M.}~\bibnamefont {Ernzerhof}},\ }\bibfield
  {title} {\bibinfo {title} {{Generalized Gradient Approximation Made
  Simple}},\ }\href {https://doi.org/10.1103/PhysRevLett.77.3865} {\bibfield
  {journal} {\bibinfo  {journal} {Phys. Rev. Lett.}\ }\textbf {\bibinfo
  {volume} {77}},\ \bibinfo {pages} {3865} (\bibinfo {year}
  {1996})}\BibitemShut {NoStop}%
\bibitem [{\citenamefont {Kawamura}(2019)}]{KAWAMURA2019197}%
  \BibitemOpen
  \bibfield  {author} {\bibinfo {author} {\bibfnamefont {M.}~\bibnamefont
  {Kawamura}},\ }\bibfield  {title} {\bibinfo {title} {{FermiSurfer:
  Fermi-surface viewer providing multiple representation schemes}},\ }\href
  {https://doi.org/https://doi.org/10.1016/j.cpc.2019.01.017} {\bibfield
  {journal} {\bibinfo  {journal} {Computer Physics Communications}\ }\textbf
  {\bibinfo {volume} {239}},\ \bibinfo {pages} {197} (\bibinfo {year}
  {2019})}\BibitemShut {NoStop}%
\bibitem [{\citenamefont {Rourke}\ and\ \citenamefont
  {Julian}(2012)}]{ROURKE2012324}%
  \BibitemOpen
  \bibfield  {author} {\bibinfo {author} {\bibfnamefont {P.}~\bibnamefont
  {Rourke}}\ and\ \bibinfo {author} {\bibfnamefont {S.}~\bibnamefont
  {Julian}},\ }\bibfield  {title} {\bibinfo {title} {{Numerical extraction of
  de Haas–van Alphen frequencies from calculated band energies}},\ }\href
  {https://doi.org/https://doi.org/10.1016/j.cpc.2011.10.015} {\bibfield
  {journal} {\bibinfo  {journal} {Computer Physics Communications}\ }\textbf
  {\bibinfo {volume} {183}},\ \bibinfo {pages} {324} (\bibinfo {year}
  {2012})}\BibitemShut {NoStop}%
\bibitem [{\citenamefont {Nie}\ \emph {et~al.}(2024)\citenamefont {Nie},
  \citenamefont {Chen}, \citenamefont {Wei}, \citenamefont {Li}, \citenamefont
  {Zhang}, \citenamefont {Mei}, \citenamefont {Wang}, \citenamefont {Song},
  \citenamefont {Song}, \citenamefont {Wang}, \citenamefont {Zhu},
  \citenamefont {Ning},\ and\ \citenamefont {Tian}}]{SrCu4xP2}%
  \BibitemOpen
  \bibfield  {author} {\bibinfo {author} {\bibfnamefont {Y.}~\bibnamefont
  {Nie}}, \bibinfo {author} {\bibfnamefont {Z.}~\bibnamefont {Chen}}, \bibinfo
  {author} {\bibfnamefont {W.}~\bibnamefont {Wei}}, \bibinfo {author}
  {\bibfnamefont {H.}~\bibnamefont {Li}}, \bibinfo {author} {\bibfnamefont
  {Y.}~\bibnamefont {Zhang}}, \bibinfo {author} {\bibfnamefont
  {M.}~\bibnamefont {Mei}}, \bibinfo {author} {\bibfnamefont {Y.}~\bibnamefont
  {Wang}}, \bibinfo {author} {\bibfnamefont {W.}~\bibnamefont {Song}}, \bibinfo
  {author} {\bibfnamefont {D.}~\bibnamefont {Song}}, \bibinfo {author}
  {\bibfnamefont {Z.}~\bibnamefont {Wang}}, \bibinfo {author} {\bibfnamefont
  {X.}~\bibnamefont {Zhu}}, \bibinfo {author} {\bibfnamefont {W.}~\bibnamefont
  {Ning}},\ and\ \bibinfo {author} {\bibfnamefont {M.}~\bibnamefont {Tian}},\
  }\bibfield  {title} {\bibinfo {title} {{Linear magnetoresistance and
  structural distortion in layered SrCu$_{4-x}$P$_2$ single crystals}},\ }\href
  {https://doi.org/10.1088/1674-1056/acf705} {\bibfield  {journal} {\bibinfo
  {journal} {Chinese Physics B}\ }\textbf {\bibinfo {volume} {33}},\ \bibinfo
  {pages} {016108} (\bibinfo {year} {2024})}\BibitemShut {NoStop}%
\bibitem [{\citenamefont {Gr\"uner}(1988)}]{CDW_Rev}%
  \BibitemOpen
  \bibfield  {author} {\bibinfo {author} {\bibfnamefont {G.}~\bibnamefont
  {Gr\"uner}},\ }\bibfield  {title} {\bibinfo {title} {{The dynamics of
  charge-density waves}},\ }\href {https://doi.org/10.1103/RevModPhys.60.1129}
  {\bibfield  {journal} {\bibinfo  {journal} {Rev. Mod. Phys.}\ }\textbf
  {\bibinfo {volume} {60}},\ \bibinfo {pages} {1129} (\bibinfo {year}
  {1988})}\BibitemShut {NoStop}%
\bibitem [{\citenamefont {Lv}\ \emph {et~al.}(2021)\citenamefont {Lv},
  \citenamefont {Qian},\ and\ \citenamefont {Ding}}]{Rev_TM}%
  \BibitemOpen
  \bibfield  {author} {\bibinfo {author} {\bibfnamefont {B.~Q.}\ \bibnamefont
  {Lv}}, \bibinfo {author} {\bibfnamefont {T.}~\bibnamefont {Qian}},\ and\
  \bibinfo {author} {\bibfnamefont {H.}~\bibnamefont {Ding}},\ }\bibfield
  {title} {\bibinfo {title} {{Experimental perspective on three-dimensional
  topological semimetals}},\ }\href
  {https://doi.org/10.1103/RevModPhys.93.025002} {\bibfield  {journal}
  {\bibinfo  {journal} {Rev. Mod. Phys.}\ }\textbf {\bibinfo {volume} {93}},\
  \bibinfo {pages} {025002} (\bibinfo {year} {2021})}\BibitemShut {NoStop}%
\bibitem [{\citenamefont {Ali}\ \emph {et~al.}(2014)\citenamefont {Ali},
  \citenamefont {Xiong}, \citenamefont {Flynn}, \citenamefont {Tao},
  \citenamefont {Gibson}, \citenamefont {Schoop}, \citenamefont {Liang},
  \citenamefont {Haldolaarachchige}, \citenamefont {Hirschberger},
  \citenamefont {Ong} \emph {et~al.}}]{WTe2}%
  \BibitemOpen
  \bibfield  {author} {\bibinfo {author} {\bibfnamefont {M.~N.}\ \bibnamefont
  {Ali}}, \bibinfo {author} {\bibfnamefont {J.}~\bibnamefont {Xiong}}, \bibinfo
  {author} {\bibfnamefont {S.}~\bibnamefont {Flynn}}, \bibinfo {author}
  {\bibfnamefont {J.}~\bibnamefont {Tao}}, \bibinfo {author} {\bibfnamefont
  {Q.~D.}\ \bibnamefont {Gibson}}, \bibinfo {author} {\bibfnamefont {L.~M.}\
  \bibnamefont {Schoop}}, \bibinfo {author} {\bibfnamefont {T.}~\bibnamefont
  {Liang}}, \bibinfo {author} {\bibfnamefont {N.}~\bibnamefont
  {Haldolaarachchige}}, \bibinfo {author} {\bibfnamefont {M.}~\bibnamefont
  {Hirschberger}}, \bibinfo {author} {\bibfnamefont {N.~P.}\ \bibnamefont
  {Ong}}, \emph {et~al.},\ }\bibfield  {title} {\bibinfo {title} {{Large,
  non-saturating magnetoresistance in WTe$_2$}},\ }\href
  {https://doi.org/https://doi.org/10.1038/nature13763} {\bibfield  {journal}
  {\bibinfo  {journal} {Nature}\ }\textbf {\bibinfo {volume} {514}},\ \bibinfo
  {pages} {205} (\bibinfo {year} {2014})}\BibitemShut {NoStop}%
\bibitem [{\citenamefont {Novak}\ \emph {et~al.}(2015)\citenamefont {Novak},
  \citenamefont {Sasaki}, \citenamefont {Segawa},\ and\ \citenamefont
  {Ando}}]{TlBiSSe}%
  \BibitemOpen
  \bibfield  {author} {\bibinfo {author} {\bibfnamefont {M.}~\bibnamefont
  {Novak}}, \bibinfo {author} {\bibfnamefont {S.}~\bibnamefont {Sasaki}},
  \bibinfo {author} {\bibfnamefont {K.}~\bibnamefont {Segawa}},\ and\ \bibinfo
  {author} {\bibfnamefont {Y.}~\bibnamefont {Ando}},\ }\bibfield  {title}
  {\bibinfo {title} {{Large linear magnetoresistance in the Dirac semimetal
  TlBiSSe}},\ }\href {https://doi.org/10.1103/PhysRevB.91.041203} {\bibfield
  {journal} {\bibinfo  {journal} {Phys. Rev. B}\ }\textbf {\bibinfo {volume}
  {91}},\ \bibinfo {pages} {041203} (\bibinfo {year} {2015})}\BibitemShut
  {NoStop}%
\bibitem [{\citenamefont {Campbell}\ \emph {et~al.}(2021)\citenamefont
  {Campbell}, \citenamefont {Collini}, \citenamefont {S{\l}awi{\'n}ska},
  \citenamefont {Autieri}, \citenamefont {Wang}, \citenamefont {Wang},
  \citenamefont {Wilfong}, \citenamefont {Eo}, \citenamefont {Neves},
  \citenamefont {Graf} \emph {et~al.}}]{FeP}%
  \BibitemOpen
  \bibfield  {author} {\bibinfo {author} {\bibfnamefont {D.}~\bibnamefont
  {Campbell}}, \bibinfo {author} {\bibfnamefont {J.}~\bibnamefont {Collini}},
  \bibinfo {author} {\bibfnamefont {J.}~\bibnamefont {S{\l}awi{\'n}ska}},
  \bibinfo {author} {\bibfnamefont {C.}~\bibnamefont {Autieri}}, \bibinfo
  {author} {\bibfnamefont {L.}~\bibnamefont {Wang}}, \bibinfo {author}
  {\bibfnamefont {K.}~\bibnamefont {Wang}}, \bibinfo {author} {\bibfnamefont
  {B.}~\bibnamefont {Wilfong}}, \bibinfo {author} {\bibfnamefont
  {Y.}~\bibnamefont {Eo}}, \bibinfo {author} {\bibfnamefont {P.}~\bibnamefont
  {Neves}}, \bibinfo {author} {\bibfnamefont {D.}~\bibnamefont {Graf}}, \emph
  {et~al.},\ }\bibfield  {title} {\bibinfo {title} {{Topologically driven
  linear magnetoresistance in helimagnetic FeP}},\ }\href
  {https://doi.org/https://doi.org/10.1038/s41535-021-00337-2} {\bibfield
  {journal} {\bibinfo  {journal} {npj Quantum Materials}\ }\textbf {\bibinfo
  {volume} {6}},\ \bibinfo {pages} {38} (\bibinfo {year} {2021})}\BibitemShut
  {NoStop}%
\bibitem [{\citenamefont {Wang}\ \emph {et~al.}(2013)\citenamefont {Wang},
  \citenamefont {Du}, \citenamefont {Xu}, \citenamefont {Zhang}, \citenamefont
  {Liu}, \citenamefont {Liu}, \citenamefont {Shi}, \citenamefont {Chen},
  \citenamefont {Wu},\ and\ \citenamefont {Zhang}}]{YPdBi}%
  \BibitemOpen
  \bibfield  {author} {\bibinfo {author} {\bibfnamefont {W.}~\bibnamefont
  {Wang}}, \bibinfo {author} {\bibfnamefont {Y.}~\bibnamefont {Du}}, \bibinfo
  {author} {\bibfnamefont {G.}~\bibnamefont {Xu}}, \bibinfo {author}
  {\bibfnamefont {X.}~\bibnamefont {Zhang}}, \bibinfo {author} {\bibfnamefont
  {E.}~\bibnamefont {Liu}}, \bibinfo {author} {\bibfnamefont {Z.}~\bibnamefont
  {Liu}}, \bibinfo {author} {\bibfnamefont {Y.}~\bibnamefont {Shi}}, \bibinfo
  {author} {\bibfnamefont {J.}~\bibnamefont {Chen}}, \bibinfo {author}
  {\bibfnamefont {G.}~\bibnamefont {Wu}},\ and\ \bibinfo {author}
  {\bibfnamefont {X.-x.}\ \bibnamefont {Zhang}},\ }\bibfield  {title} {\bibinfo
  {title} {{Large linear magnetoresistance and Shubnikov-de Hass oscillations
  in single crystals of YPdBi Heusler topological insulators}},\ }\href
  {https://doi.org/https://doi.org/10.1038/srep02181} {\bibfield  {journal}
  {\bibinfo  {journal} {Scientific reports}\ }\textbf {\bibinfo {volume} {3}},\
  \bibinfo {pages} {2181} (\bibinfo {year} {2013})}\BibitemShut {NoStop}%
\bibitem [{\citenamefont {Shiomi}\ and\ \citenamefont {Saitoh}(2017)}]{Ru2Sn3}%
  \BibitemOpen
  \bibfield  {author} {\bibinfo {author} {\bibfnamefont {Y.}~\bibnamefont
  {Shiomi}}\ and\ \bibinfo {author} {\bibfnamefont {E.}~\bibnamefont
  {Saitoh}},\ }\bibfield  {title} {\bibinfo {title} {{Linear magnetoresistance
  in a topological insulator Ru$_2$Sn$_3$}},\ }\href
  {https://doi.org/10.1063/1.4978773} {\bibfield  {journal} {\bibinfo
  {journal} {AIP Advances}\ }\textbf {\bibinfo {volume} {7}},\ \bibinfo {pages}
  {035011} (\bibinfo {year} {2017})}\BibitemShut {NoStop}%
\bibitem [{\citenamefont {Abrikosov}(1998)}]{Abrikosov}%
  \BibitemOpen
  \bibfield  {author} {\bibinfo {author} {\bibfnamefont {A.~A.}\ \bibnamefont
  {Abrikosov}},\ }\bibfield  {title} {\bibinfo {title} {{Quantum
  magnetoresistance}},\ }\href {https://doi.org/10.1103/PhysRevB.58.2788}
  {\bibfield  {journal} {\bibinfo  {journal} {Phys. Rev. B}\ }\textbf {\bibinfo
  {volume} {58}},\ \bibinfo {pages} {2788} (\bibinfo {year}
  {1998})}\BibitemShut {NoStop}%
\bibitem [{\citenamefont {Hu}\ \emph {et~al.}(2019)\citenamefont {Hu},
  \citenamefont {Xu}, \citenamefont {Ni},\ and\ \citenamefont
  {Mao}}]{Tranport_topo}%
  \BibitemOpen
  \bibfield  {author} {\bibinfo {author} {\bibfnamefont {J.}~\bibnamefont
  {Hu}}, \bibinfo {author} {\bibfnamefont {S.-Y.}\ \bibnamefont {Xu}}, \bibinfo
  {author} {\bibfnamefont {N.}~\bibnamefont {Ni}},\ and\ \bibinfo {author}
  {\bibfnamefont {Z.}~\bibnamefont {Mao}},\ }\bibfield  {title} {\bibinfo
  {title} {Transport of topological semimetals},\ }\href
  {https://doi.org/https://doi.org/10.1146/annurev-matsci-070218-010023}
  {\bibfield  {journal} {\bibinfo  {journal} {Annual Review of Materials
  Research}\ }\textbf {\bibinfo {volume} {49}},\ \bibinfo {pages} {207}
  (\bibinfo {year} {2019})}\BibitemShut {NoStop}%
\bibitem [{\citenamefont {Zeng}\ \emph {et~al.}(2016)\citenamefont {Zeng},
  \citenamefont {Lou}, \citenamefont {Wu}, \citenamefont {Xu}, \citenamefont
  {Guo}, \citenamefont {Kong}, \citenamefont {Zhong}, \citenamefont {Ma},
  \citenamefont {Fu}, \citenamefont {Richard}, \citenamefont {Wang},
  \citenamefont {Liu}, \citenamefont {Lu}, \citenamefont {Huang}, \citenamefont
  {Fang}, \citenamefont {Sun}, \citenamefont {Wang}, \citenamefont {Wang},
  \citenamefont {Shi}, \citenamefont {Weng}, \citenamefont {Lei}, \citenamefont
  {Liu}, \citenamefont {Wang}, \citenamefont {Qian}, \citenamefont {Luo},\ and\
  \citenamefont {Ding}}]{LaSb}%
  \BibitemOpen
  \bibfield  {author} {\bibinfo {author} {\bibfnamefont {L.-K.}\ \bibnamefont
  {Zeng}}, \bibinfo {author} {\bibfnamefont {R.}~\bibnamefont {Lou}}, \bibinfo
  {author} {\bibfnamefont {D.-S.}\ \bibnamefont {Wu}}, \bibinfo {author}
  {\bibfnamefont {Q.~N.}\ \bibnamefont {Xu}}, \bibinfo {author} {\bibfnamefont
  {P.-J.}\ \bibnamefont {Guo}}, \bibinfo {author} {\bibfnamefont {L.-Y.}\
  \bibnamefont {Kong}}, \bibinfo {author} {\bibfnamefont {Y.-G.}\ \bibnamefont
  {Zhong}}, \bibinfo {author} {\bibfnamefont {J.-Z.}\ \bibnamefont {Ma}},
  \bibinfo {author} {\bibfnamefont {B.-B.}\ \bibnamefont {Fu}}, \bibinfo
  {author} {\bibfnamefont {P.}~\bibnamefont {Richard}}, \bibinfo {author}
  {\bibfnamefont {P.}~\bibnamefont {Wang}}, \bibinfo {author} {\bibfnamefont
  {G.~T.}\ \bibnamefont {Liu}}, \bibinfo {author} {\bibfnamefont
  {L.}~\bibnamefont {Lu}}, \bibinfo {author} {\bibfnamefont {Y.-B.}\
  \bibnamefont {Huang}}, \bibinfo {author} {\bibfnamefont {C.}~\bibnamefont
  {Fang}}, \bibinfo {author} {\bibfnamefont {S.-S.}\ \bibnamefont {Sun}},
  \bibinfo {author} {\bibfnamefont {Q.}~\bibnamefont {Wang}}, \bibinfo {author}
  {\bibfnamefont {L.}~\bibnamefont {Wang}}, \bibinfo {author} {\bibfnamefont
  {Y.-G.}\ \bibnamefont {Shi}}, \bibinfo {author} {\bibfnamefont {H.~M.}\
  \bibnamefont {Weng}}, \bibinfo {author} {\bibfnamefont {H.-C.}\ \bibnamefont
  {Lei}}, \bibinfo {author} {\bibfnamefont {K.}~\bibnamefont {Liu}}, \bibinfo
  {author} {\bibfnamefont {S.-C.}\ \bibnamefont {Wang}}, \bibinfo {author}
  {\bibfnamefont {T.}~\bibnamefont {Qian}}, \bibinfo {author} {\bibfnamefont
  {J.-L.}\ \bibnamefont {Luo}},\ and\ \bibinfo {author} {\bibfnamefont
  {H.}~\bibnamefont {Ding}},\ }\bibfield  {title} {\bibinfo {title}
  {{Compensated Semimetal LaSb with Unsaturated Magnetoresistance}},\ }\href
  {https://doi.org/10.1103/PhysRevLett.117.127204} {\bibfield  {journal}
  {\bibinfo  {journal} {Phys. Rev. Lett.}\ }\textbf {\bibinfo {volume} {117}},\
  \bibinfo {pages} {127204} (\bibinfo {year} {2016})}\BibitemShut {NoStop}%
\bibitem [{\citenamefont {Yuan}\ \emph {et~al.}(2016)\citenamefont {Yuan},
  \citenamefont {Lu}, \citenamefont {Liu}, \citenamefont {Wang},\ and\
  \citenamefont {Jia}}]{TaAs2}%
  \BibitemOpen
  \bibfield  {author} {\bibinfo {author} {\bibfnamefont {Z.}~\bibnamefont
  {Yuan}}, \bibinfo {author} {\bibfnamefont {H.}~\bibnamefont {Lu}}, \bibinfo
  {author} {\bibfnamefont {Y.}~\bibnamefont {Liu}}, \bibinfo {author}
  {\bibfnamefont {J.}~\bibnamefont {Wang}},\ and\ \bibinfo {author}
  {\bibfnamefont {S.}~\bibnamefont {Jia}},\ }\bibfield  {title} {\bibinfo
  {title} {{Large magnetoresistance in compensated semimetals
  ${\mathrm{TaAs}}_{2}$ and ${\mathrm{NbAs}}_{2}$}},\ }\href
  {https://doi.org/10.1103/PhysRevB.93.184405} {\bibfield  {journal} {\bibinfo
  {journal} {Phys. Rev. B}\ }\textbf {\bibinfo {volume} {93}},\ \bibinfo
  {pages} {184405} (\bibinfo {year} {2016})}\BibitemShut {NoStop}%
\bibitem [{\citenamefont {Parish}\ and\ \citenamefont
  {Littlewood}(2003)}]{Ag2Se}%
  \BibitemOpen
  \bibfield  {author} {\bibinfo {author} {\bibfnamefont {M.}~\bibnamefont
  {Parish}}\ and\ \bibinfo {author} {\bibfnamefont {P.}~\bibnamefont
  {Littlewood}},\ }\bibfield  {title} {\bibinfo {title} {{Non-saturating
  magnetoresistance in heavily disordered semiconductors}},\ }\href
  {https://doi.org/https://doi.org/10.1038/nature02073} {\bibfield  {journal}
  {\bibinfo  {journal} {Nature}\ }\textbf {\bibinfo {volume} {426}},\ \bibinfo
  {pages} {162} (\bibinfo {year} {2003})}\BibitemShut {NoStop}%
\bibitem [{\citenamefont {Liang}\ \emph {et~al.}(2015)\citenamefont {Liang},
  \citenamefont {Gibson}, \citenamefont {Ali}, \citenamefont {Liu},
  \citenamefont {Cava},\ and\ \citenamefont {Ong}}]{Cd3As2}%
  \BibitemOpen
  \bibfield  {author} {\bibinfo {author} {\bibfnamefont {T.}~\bibnamefont
  {Liang}}, \bibinfo {author} {\bibfnamefont {Q.}~\bibnamefont {Gibson}},
  \bibinfo {author} {\bibfnamefont {M.~N.}\ \bibnamefont {Ali}}, \bibinfo
  {author} {\bibfnamefont {M.}~\bibnamefont {Liu}}, \bibinfo {author}
  {\bibfnamefont {R.}~\bibnamefont {Cava}},\ and\ \bibinfo {author}
  {\bibfnamefont {N.}~\bibnamefont {Ong}},\ }\bibfield  {title} {\bibinfo
  {title} {{Ultrahigh mobility and giant magnetoresistance in the Dirac
  semimetal Cd$_3$As$_2$}},\ }\href
  {https://doi.org/https://doi.org/10.1038/nmat4143} {\bibfield  {journal}
  {\bibinfo  {journal} {Nature materials}\ }\textbf {\bibinfo {volume} {14}},\
  \bibinfo {pages} {280} (\bibinfo {year} {2015})}\BibitemShut {NoStop}%
\bibitem [{\citenamefont {Narayanan}\ \emph {et~al.}(2015)\citenamefont
  {Narayanan}, \citenamefont {Watson}, \citenamefont {Blake}, \citenamefont
  {Bruyant}, \citenamefont {Drigo}, \citenamefont {Chen}, \citenamefont
  {Prabhakaran}, \citenamefont {Yan}, \citenamefont {Felser}, \citenamefont
  {Kong}, \citenamefont {Canfield},\ and\ \citenamefont {Coldea}}]{n-Cd3As2}%
  \BibitemOpen
  \bibfield  {author} {\bibinfo {author} {\bibfnamefont {A.}~\bibnamefont
  {Narayanan}}, \bibinfo {author} {\bibfnamefont {M.~D.}\ \bibnamefont
  {Watson}}, \bibinfo {author} {\bibfnamefont {S.~F.}\ \bibnamefont {Blake}},
  \bibinfo {author} {\bibfnamefont {N.}~\bibnamefont {Bruyant}}, \bibinfo
  {author} {\bibfnamefont {L.}~\bibnamefont {Drigo}}, \bibinfo {author}
  {\bibfnamefont {Y.~L.}\ \bibnamefont {Chen}}, \bibinfo {author}
  {\bibfnamefont {D.}~\bibnamefont {Prabhakaran}}, \bibinfo {author}
  {\bibfnamefont {B.}~\bibnamefont {Yan}}, \bibinfo {author} {\bibfnamefont
  {C.}~\bibnamefont {Felser}}, \bibinfo {author} {\bibfnamefont
  {T.}~\bibnamefont {Kong}}, \bibinfo {author} {\bibfnamefont {P.~C.}\
  \bibnamefont {Canfield}},\ and\ \bibinfo {author} {\bibfnamefont {A.~I.}\
  \bibnamefont {Coldea}},\ }\bibfield  {title} {\bibinfo {title} {Linear
  magnetoresistance caused by mobility fluctuations in $n$-doped
  cd$_3$as$_2$},\ }\href {https://doi.org/10.1103/PhysRevLett.114.117201}
  {\bibfield  {journal} {\bibinfo  {journal} {Phys. Rev. Lett.}\ }\textbf
  {\bibinfo {volume} {114}},\ \bibinfo {pages} {117201} (\bibinfo {year}
  {2015})}\BibitemShut {NoStop}%
\bibitem [{\citenamefont {Laha}\ \emph {et~al.}(2019)\citenamefont {Laha},
  \citenamefont {Malick}, \citenamefont {Singha}, \citenamefont {Mandal},
  \citenamefont {Rambabu}, \citenamefont {Kanchana},\ and\ \citenamefont
  {Hossain}}]{YbCdGe}%
  \BibitemOpen
  \bibfield  {author} {\bibinfo {author} {\bibfnamefont {A.}~\bibnamefont
  {Laha}}, \bibinfo {author} {\bibfnamefont {S.}~\bibnamefont {Malick}},
  \bibinfo {author} {\bibfnamefont {R.}~\bibnamefont {Singha}}, \bibinfo
  {author} {\bibfnamefont {P.}~\bibnamefont {Mandal}}, \bibinfo {author}
  {\bibfnamefont {P.}~\bibnamefont {Rambabu}}, \bibinfo {author} {\bibfnamefont
  {V.}~\bibnamefont {Kanchana}},\ and\ \bibinfo {author} {\bibfnamefont
  {Z.}~\bibnamefont {Hossain}},\ }\bibfield  {title} {\bibinfo {title}
  {{Magnetotransport properties of the correlated topological nodal-line
  semimetal YbCdGe}},\ }\href {https://doi.org/10.1103/PhysRevB.99.241102}
  {\bibfield  {journal} {\bibinfo  {journal} {Phys. Rev. B}\ }\textbf {\bibinfo
  {volume} {99}},\ \bibinfo {pages} {241102} (\bibinfo {year}
  {2019})}\BibitemShut {NoStop}%
\bibitem [{\citenamefont {Hu}\ \emph {et~al.}(2016)\citenamefont {Hu},
  \citenamefont {Tang}, \citenamefont {Liu}, \citenamefont {Liu}, \citenamefont
  {Zhu}, \citenamefont {Graf}, \citenamefont {Myhro}, \citenamefont {Tran},
  \citenamefont {Lau}, \citenamefont {Wei},\ and\ \citenamefont
  {Mao}}]{ZrSiSe}%
  \BibitemOpen
  \bibfield  {author} {\bibinfo {author} {\bibfnamefont {J.}~\bibnamefont
  {Hu}}, \bibinfo {author} {\bibfnamefont {Z.}~\bibnamefont {Tang}}, \bibinfo
  {author} {\bibfnamefont {J.}~\bibnamefont {Liu}}, \bibinfo {author}
  {\bibfnamefont {X.}~\bibnamefont {Liu}}, \bibinfo {author} {\bibfnamefont
  {Y.}~\bibnamefont {Zhu}}, \bibinfo {author} {\bibfnamefont {D.}~\bibnamefont
  {Graf}}, \bibinfo {author} {\bibfnamefont {K.}~\bibnamefont {Myhro}},
  \bibinfo {author} {\bibfnamefont {S.}~\bibnamefont {Tran}}, \bibinfo {author}
  {\bibfnamefont {C.~N.}\ \bibnamefont {Lau}}, \bibinfo {author} {\bibfnamefont
  {J.}~\bibnamefont {Wei}},\ and\ \bibinfo {author} {\bibfnamefont
  {Z.}~\bibnamefont {Mao}},\ }\bibfield  {title} {\bibinfo {title} {{Evidence
  of Topological Nodal-Line Fermions in ZrSiSe and ZrSiTe}},\ }\href
  {https://doi.org/10.1103/PhysRevLett.117.016602} {\bibfield  {journal}
  {\bibinfo  {journal} {Phys. Rev. Lett.}\ }\textbf {\bibinfo {volume} {117}},\
  \bibinfo {pages} {016602} (\bibinfo {year} {2016})}\BibitemShut {NoStop}%
\bibitem [{\citenamefont {Zhang}\ \emph {et~al.}(2017)\citenamefont {Zhang},
  \citenamefont {Gao}, \citenamefont {Zhang}, \citenamefont {Wang},
  \citenamefont {Zhang}, \citenamefont {Zhang}, \citenamefont {Niu},
  \citenamefont {Zhang},\ and\ \citenamefont {Xu}}]{ZrSiS}%
  \BibitemOpen
  \bibfield  {author} {\bibinfo {author} {\bibfnamefont {J.}~\bibnamefont
  {Zhang}}, \bibinfo {author} {\bibfnamefont {M.}~\bibnamefont {Gao}}, \bibinfo
  {author} {\bibfnamefont {J.}~\bibnamefont {Zhang}}, \bibinfo {author}
  {\bibfnamefont {X.}~\bibnamefont {Wang}}, \bibinfo {author} {\bibfnamefont
  {X.}~\bibnamefont {Zhang}}, \bibinfo {author} {\bibfnamefont
  {M.}~\bibnamefont {Zhang}}, \bibinfo {author} {\bibfnamefont
  {W.}~\bibnamefont {Niu}}, \bibinfo {author} {\bibfnamefont {R.}~\bibnamefont
  {Zhang}},\ and\ \bibinfo {author} {\bibfnamefont {Y.}~\bibnamefont {Xu}},\
  }\bibfield  {title} {\bibinfo {title} {{Transport evidence of 3D topological
  nodal-line semimetal phase in ZrSiS}},\ }\href
  {https://doi.org/10.1007/s11467-017-0705-7} {\bibfield  {journal} {\bibinfo
  {journal} {Frontiers of Physics}\ }\textbf {\bibinfo {volume} {13}},\
  \bibinfo {pages} {137201} (\bibinfo {year} {2017})}\BibitemShut {NoStop}%
\bibitem [{\citenamefont {Shoenberg}(2009)}]{Shoenberg}%
  \BibitemOpen
  \bibfield  {author} {\bibinfo {author} {\bibfnamefont {D.}~\bibnamefont
  {Shoenberg}},\ }\href@noop {} {\emph {\bibinfo {title} {{Magnetic
  oscillations in metals}}}}\ (\bibinfo  {publisher} {Cambridge university
  press, Cambridge, UK},\ \bibinfo {year} {2009})\BibitemShut {NoStop}%
\bibitem [{\citenamefont {Ortiz}\ \emph {et~al.}(2019)\citenamefont {Ortiz},
  \citenamefont {Gomes}, \citenamefont {Morey}, \citenamefont {Winiarski},
  \citenamefont {Bordelon}, \citenamefont {Mangum}, \citenamefont {Oswald},
  \citenamefont {Rodriguez-Rivera}, \citenamefont {Neilson}, \citenamefont
  {Wilson}, \citenamefont {Ertekin}, \citenamefont {McQueen},\ and\
  \citenamefont {Toberer}}]{Ortiz2019}%
  \BibitemOpen
  \bibfield  {author} {\bibinfo {author} {\bibfnamefont {B.~R.}\ \bibnamefont
  {Ortiz}}, \bibinfo {author} {\bibfnamefont {L.~C.}\ \bibnamefont {Gomes}},
  \bibinfo {author} {\bibfnamefont {J.~R.}\ \bibnamefont {Morey}}, \bibinfo
  {author} {\bibfnamefont {M.}~\bibnamefont {Winiarski}}, \bibinfo {author}
  {\bibfnamefont {M.}~\bibnamefont {Bordelon}}, \bibinfo {author}
  {\bibfnamefont {J.~S.}\ \bibnamefont {Mangum}}, \bibinfo {author}
  {\bibfnamefont {I.~W.~H.}\ \bibnamefont {Oswald}}, \bibinfo {author}
  {\bibfnamefont {J.~A.}\ \bibnamefont {Rodriguez-Rivera}}, \bibinfo {author}
  {\bibfnamefont {J.~R.}\ \bibnamefont {Neilson}}, \bibinfo {author}
  {\bibfnamefont {S.~D.}\ \bibnamefont {Wilson}}, \bibinfo {author}
  {\bibfnamefont {E.}~\bibnamefont {Ertekin}}, \bibinfo {author} {\bibfnamefont
  {T.~M.}\ \bibnamefont {McQueen}},\ and\ \bibinfo {author} {\bibfnamefont
  {E.~S.}\ \bibnamefont {Toberer}},\ }\bibfield  {title} {\bibinfo {title}
  {{New kagome prototype materials: discovery of \ce{KV3Sb5}, \ce{RbV3Sb5} and
  \ce{CsV3Sb5}}},\ }\href {https://doi.org/10.1103/PhysRevMaterials.3.094407}
  {\bibfield  {journal} {\bibinfo  {journal} {Phys. Rev. Mater.}\ }\textbf
  {\bibinfo {volume} {3}},\ \bibinfo {pages} {094407} (\bibinfo {year}
  {2019})}\BibitemShut {NoStop}%
\bibitem [{\citenamefont {Górnicka}\ \emph {et~al.}(2021)\citenamefont
  {Górnicka}, \citenamefont {Gui}, \citenamefont {Wiendlocha}, \citenamefont
  {Nguyen}, \citenamefont {Xie}, \citenamefont {Cava},\ and\ \citenamefont
  {Klimczuk}}]{Gornicka2021}%
  \BibitemOpen
  \bibfield  {author} {\bibinfo {author} {\bibfnamefont {K.}~\bibnamefont
  {Górnicka}}, \bibinfo {author} {\bibfnamefont {X.}~\bibnamefont {Gui}},
  \bibinfo {author} {\bibfnamefont {B.}~\bibnamefont {Wiendlocha}}, \bibinfo
  {author} {\bibfnamefont {L.~T.}\ \bibnamefont {Nguyen}}, \bibinfo {author}
  {\bibfnamefont {W.}~\bibnamefont {Xie}}, \bibinfo {author} {\bibfnamefont
  {R.~J.}\ \bibnamefont {Cava}},\ and\ \bibinfo {author} {\bibfnamefont
  {T.}~\bibnamefont {Klimczuk}},\ }\bibfield  {title} {\bibinfo {title}
  {{\ce{NbIr2B2} and \ce{TaIr2B2} – New Low Symmetry Noncentrosymmetric
  Superconductors with Strong Spin–Orbit Coupling}},\ }\href
  {https://doi.org/https://doi.org/10.1002/adfm.202007960} {\bibfield
  {journal} {\bibinfo  {journal} {Adv. Funct. Mater.}\ }\textbf {\bibinfo
  {volume} {31}},\ \bibinfo {pages} {2007960} (\bibinfo {year}
  {2021})}\BibitemShut {NoStop}%
\bibitem [{\citenamefont {Lygouras}\ \emph {et~al.}(2025)\citenamefont
  {Lygouras}, \citenamefont {Zhang}, \citenamefont {Gautreau}, \citenamefont
  {Pula}, \citenamefont {Sharma}, \citenamefont {Gao}, \citenamefont {Berry},
  \citenamefont {Halloran}, \citenamefont {Orban}, \citenamefont
  {Grissonnanche}, \citenamefont {Chamorro}, \citenamefont {Mikuri},
  \citenamefont {Bhoi}, \citenamefont {Siegler}, \citenamefont {Livi},
  \citenamefont {Uwatoko}, \citenamefont {Nakatsuji}, \citenamefont {Ramshaw},
  \citenamefont {Li}, \citenamefont {Luke}, \citenamefont {Broholm},\ and\
  \citenamefont {McQueen}}]{Lygouras2025}%
  \BibitemOpen
  \bibfield  {author} {\bibinfo {author} {\bibfnamefont {C.~J.}\ \bibnamefont
  {Lygouras}}, \bibinfo {author} {\bibfnamefont {J.}~\bibnamefont {Zhang}},
  \bibinfo {author} {\bibfnamefont {J.}~\bibnamefont {Gautreau}}, \bibinfo
  {author} {\bibfnamefont {M.}~\bibnamefont {Pula}}, \bibinfo {author}
  {\bibfnamefont {S.}~\bibnamefont {Sharma}}, \bibinfo {author} {\bibfnamefont
  {S.}~\bibnamefont {Gao}}, \bibinfo {author} {\bibfnamefont {T.}~\bibnamefont
  {Berry}}, \bibinfo {author} {\bibfnamefont {T.}~\bibnamefont {Halloran}},
  \bibinfo {author} {\bibfnamefont {P.}~\bibnamefont {Orban}}, \bibinfo
  {author} {\bibfnamefont {G.}~\bibnamefont {Grissonnanche}}, \bibinfo {author}
  {\bibfnamefont {J.~R.}\ \bibnamefont {Chamorro}}, \bibinfo {author}
  {\bibfnamefont {T.}~\bibnamefont {Mikuri}}, \bibinfo {author} {\bibfnamefont
  {D.~K.}\ \bibnamefont {Bhoi}}, \bibinfo {author} {\bibfnamefont {M.~A.}\
  \bibnamefont {Siegler}}, \bibinfo {author} {\bibfnamefont {K.~J.}\
  \bibnamefont {Livi}}, \bibinfo {author} {\bibfnamefont {Y.}~\bibnamefont
  {Uwatoko}}, \bibinfo {author} {\bibfnamefont {S.}~\bibnamefont {Nakatsuji}},
  \bibinfo {author} {\bibfnamefont {B.~J.}\ \bibnamefont {Ramshaw}}, \bibinfo
  {author} {\bibfnamefont {Y.}~\bibnamefont {Li}}, \bibinfo {author}
  {\bibfnamefont {G.~M.}\ \bibnamefont {Luke}}, \bibinfo {author}
  {\bibfnamefont {C.~L.}\ \bibnamefont {Broholm}},\ and\ \bibinfo {author}
  {\bibfnamefont {T.~M.}\ \bibnamefont {McQueen}},\ }\bibfield  {title}
  {\bibinfo {title} {{Type {I} and type {II} superconductivity in a quasi-2D
  Dirac metal}},\ }\href {https://doi.org/10.1039/D5MA00022J} {\bibfield
  {journal} {\bibinfo  {journal} {Mater. Adv.}\ }\textbf {\bibinfo {volume}
  {6}},\ \bibinfo {pages} {1685} (\bibinfo {year} {2025})}\BibitemShut
  {NoStop}%
\end{thebibliography}%
	
\end{document}